\documentclass[reprint,aps,prd,nofootinbib,superscriptaddress,floatfix]{revtex4-2}
\usepackage[utf8]{inputenc}
\usepackage{amsmath}
\usepackage{amsthm}
\usepackage{natbib}
\usepackage{amsfonts}
\usepackage{microtype}
\usepackage{hyperref}
\usepackage{cleveref}
\usepackage{amssymb}
\usepackage{graphicx}
\usepackage[usenames,dvipsnames]{color}
\usepackage{mathrsfs}
\usepackage{orcidlink}
\usepackage{upgreek}

\allowdisplaybreaks

\def\lsim{~\rlap{$<$}{\lower 1.0ex\hbox{$\sim$}}}
\def\bsim{~\rlap{$>$}{\lower 1.0ex\hbox{$\sim$}}}

\def\gpc{\ {\rm Gpc}}

\def\msun{\ {\rm M_\odot}}

\def\AU{\ {\rm AU}}

\def\Hz{\ {\rm Hz}}

\def\yr{\ {\rm yr}}

\def\apjl{Astrophys.\ J.\ Lett.}
\def\mnras{Mon.\ Not.\ R.\ Astron.\ Soc.}

\def\aap{Astron.\ Astrophys.}
\def\apj{Astrophys.\ J.}

\def\apjs{Astrophys.\ J. Supp.}
\def\prl{Phys.\ Rev.\ Lett.}
\def\prd{Phys.\ Rev.\ D}

\def\jcap{{JCAP\ }}

\theoremstyle{definition}

\theoremstyle{remark}

\numberwithin{equation}{section}
\numberwithin{thm}{section}

\newcommand{\abs}[1]{\left\vert#1\right\vert}
\newcommand{\set}[1]{\left\{#1\right\}}

\newcommand{\be}{\begin{equation}}
\newcommand{\ee}{\end{equation}}

\def\dd{\mathrm{d}}

\let\ln\relax
\DeclareMathOperator{\ln}{ln}

\def\mathbi#1{\textbf{\em #1}}

\renewcommand{\v}[1]{\bm{#1}}

\def\vxi{\boldsymbol{\xi}}

\def\nvh{\hat{\mathbi{n}}}

\def\vr{\mathbi{r}}

\newcommand{\xhat}{\mathbf{\hat{x}}}

\newcommand{\ret}{\textrm{ret}}

\def\fmin{f_\text{min}}
\def\fmax{f_\text{max}}
\def\fisco{f_\text{\tiny ISCO}}

\def\ogw{\Omega_\text{gw}}

\def\fmin{f_\text{min}}
\def\fmax{f_\text{max}}

\begin{document}
	
\title{Frequency-Domain Distribution of Astrophysical Gravitational-Wave Backgrounds}
	
\author{Yonadav Barry Ginat\,\orcidlink{0000-0003-1992-1910}}
\email{yb.ginat@physics.ox.ac.uk}
\affiliation{Rudolf Peierls Centre for Theoretical Physics, University of Oxford, Parks Road, Oxford, OX1 3PU, United Kingdom}%
\affiliation{New College, Holywell Street, Oxford, OX1 3BN, United Kingdom}%
\affiliation{Physics department, Technion, Haifa 3200003, Israel}
\author{Robert Reischke\,\orcidlink{0000-0001-5404-8753}}
\email{reischke@astro.ruhr-uni-bochum.de}
\affiliation{Ruhr University Bochum, Faculty of Physics and Astronomy, Astronomical Institute (AIRUB), German Centre for Cosmological Lensing, 44780 Bochum, Germany}
\affiliation{Argelander-Institut für Astronomie, Universität Bonn, Auf dem Hügel 71, D-53121 Bonn, Germany}
\author{Ivan Rapoport\,\orcidlink{0009-0007-7887-783X}}
\email{ivanr@campus.technion.ac.il}
\affiliation{Physics department, Technion, Haifa 3200003, Israel}
\author{Vincent Desjacques\,\orcidlink{0000-0003-2062-8172}}
\email{dvince@physics.technion.ac.il}
\affiliation{Physics department, Technion, Haifa 3200003, Israel}

\begin{abstract}
    The superposition of many astrophysical gravitational wave (GW) signals below typical detection thresholds baths detectors in a stochastic gravitational wave background (SGWB).
    In this work, we present a Fourier space approach to compute the frequency-domain distribution of stochastic gravitational wave backgrounds produced by discrete sources. 
    Expressions for the moment-generating function and the distribution of observed (discrete) Fourier modes are provided. The results are first applied to the signal originating from all the mergers of compact stellar remnants (black holes and neutron stars) in the Universe, which is found to exhibit a $-4$ power-law tail. This tail is verified in the signal-to-noise ratio distribution of GWTC events. 
    The extent to which the subtraction of bright (loud) mergers gaussianizes the resulting confusion noise of unresolved sources is then illustrated. 
    The power-law asymptotic tail for the unsubtracted signal, and an exponentially decaying tail in the case of the SGWB, are also derived analytically. Our results generalize to any background of gravitational waves emanating from discrete, individually coherent, sources.
\end{abstract}

\maketitle

\section{Introduction}
\label{sec:introduction}

The recent direct detections of gravitational waves (GWs) from a binary black hole (BH) mergers \cite{LIGOVirgo2016,LIGOvirgo2019PRX,LIGOvirgo2021PRX} have opened a new window to probe cosmic structure formation and evolution. However, the weakness of gravity implies that the amplitude of gravitational waves is generally small. Therefore, unlike the ``bright'' (loud) binary BH mergers detected so far, many astrophysical GW sources will not be detected by forthcoming experiments. The cumulative effect of a large number of unresolved astrophysical GW sources on our past light-cone results in stochastic gravitational wave backgrounds (SGWBs) and may, when investigated, reveal details of their physical origin \citep[e.g.][and references therein]{michelson:1987,christensen:1992,flanagan:1993,allen/romano:1997,KosenkoPostnov2000,phinney:2001,schneider/ferrari/etal:2001,Coward:2002ba,FarmerPhinney2003,Timpanoetal2006,Regimbau:2009rk,rosado:2011,Regimbau:2011rp,zhu/howell/etal:2011,wu/mandic/regimbau:2012,Callisteretal2016,romano/cornish:2017,Brito:2017wnc,Jenkinsetal2018,Barausse:2018vdb,renzini/contaldi:2018,caprini/figueroa:2018,Conneelyetal2019,Bartolo:2019oiq,Chang:2021afa,renzini/etal:2022,Banks:2023eym}. The current upper limit on the energy density of the SGWB produced by mergers of compact stellar remnants in the Universe is $\ogw\leq 3.4\times 10^{-9}$ at $f=25\Hz$ (assuming a power-law background of spectral index 2/3), derived from {\small LIGO}-Virgo-{\small KAGRA}'s O3 run \cite{LIGOVIRGOSGWB}. Future experiments such as {\small LISA} or the Einstein Telescope \citep{LISApaper1,LISApaper2,ETpaper1,ETpaper2} should probe this background along with other SGWBs of cosmological origin.

SGWBs of cosmological origin (such as the primordial GWs produced by quantum fluctuations during inflation) are nearly Gaussian random fields due to the random nature of the sources \cite[][for reviews]{Maggiore2,caprini/figueroa:2018}. However, the situation is more complex for signals of astrophysical origin \citep[see the discussion in][]{romano/cornish:2017} because wave-forms produced by astrophysical sources such as compact binaries are purely deterministic and coherent while the source properties (masses, separation etc.) and spatial distribution are intrinsically stochastic. In practice, bright sources can be identified and subtracted out of the signal (down to a threshold which depends on the sensitivity of the detector), leaving behind a residual non-deterministic confusion noise, known as the SGWB. The statistical properties of the residual background depend on the number of superposed signals. In particular, these are given by the number of active sources $N_0$, the duration $T$ of the experiment and the details of the bright source subtraction. The Central Limit Theorem (CLT) guarantees that the distribution of the unsubtracted signal and the resulting SGWB converges towards a Gaussian when $N_0$ and/or $T$ tends to infinity. In this case, knowledge of the second moment -- the power spectrum -- suffices to determine all higher-order moments. For non-Gaussian signals, information is also encoded in higher-order moments.

In this work, we quantify the statistical properties of these signals in frequency space using a rigorous approach similar to that used, e.g., for the large-scale structure of the Universe \citep[see, e.g.,][]{matsubara:2003,matsubara:2007}; and thereby expand the frequency-domain studies of e.g. \cite{rosado:2011,Meacher:2014aca,Meacher:2015iua, romano/cornish:2017} and the time-domain analysis of \cite{Ginatetal2020}. This allows us to characterize precisely the distribution of the SGWB as a function of detector sensitivity, the bright source subtraction method etc. without resorting to Monte-Carlo simulations.

The paper is organized as follows: In \S\ref{sec:theory}, we spell out our approach to calculate the distribution of the signal in frequency domain. We provide a general expression for the moment-generating function, the distribution of observed (discrete) Fourier modes and their regularization, and derive its large-strain asymptotic expansion analytically.
We apply our approach to the unsubtracted signal produced by mergers of compact stellar remnants in the Universe. The short presentation of the physical model in \S\ref{sec:compact mergers} is followed by a detailed presentation of our results in \S\ref{sec:results for probabilities}. After we demonstrate the consistency of our approach with previous literature at the power spectrum level, we compute the frequency-domain distribution of the unsubtracted signal, and the resulting SGWB\footnote{In \cite{Ginatetal2020} this is referred to as the confusion background.} obtained after the subtraction of bright mergers. We conclude in \S\ref{sec:conclusion}. A flat $\Lambda$CDM cosmology will be assumed throughout this paper \cite{Planck2018}.

\section{Fourier analysis of discrete Stochastic GW Signals}
\label{sec:theory}

We refer the reader to \cite{Maggiore} for a textbook reference on gravitational waves. Here and henceforth, $f$ and $\nvh$ will denote the GW frequency and sky direction in the detector's frame. We will also assume that the GW signal is stationary for realistic observation times $T\ll t_0, H_0^{-1}$, where $t_0$ is the age of the Universe, and $H_0$ is the present-day Hubble constant. This property alone already implies that strains at different frequencies are uncorrelated.

\subsection{Fourier modes}
\label{sec:fourier}

In this sub-section, we set out the Fourier transform conventions used in this paper.
For a GW detector located at the origin of the coordinate system chosen here, the GW strain produced by $N_0$ discrete sources can be generally decomposed into (this expression defines our Fourier convention)
\be
h_{ij}(t) = \sum_{\alpha=1}^{N_0} \sum_{A=+,\times} \int_{-\infty}^{+\infty}\!\!\dd f  ~\tilde h_{\alpha A}(f,\nvh_\alpha)\,e_{ij}^A(\nvh_\alpha)\, \mathrm{e}^{-2\pi \mathrm{i} f t}
\ee
where $\nvh_\alpha$ is the propagation direction of the GW signal from source $\alpha$, and the corresponding Fourier modes $\tilde h_{\alpha A}(f,\nvh_\alpha)$ (defined for all frequencies $-\infty<f<+\infty$) depend on the polarization $A$. The reality condition also implies $\tilde h_{\alpha A}(-f,\nvh_\alpha)=\tilde h_{\alpha A}^*(f,\nvh_\alpha)$, so we just focus on positive frequencies.

The continuous (scalar) output of the detector is of the form $h(t)=D^{ij}h_{ij}$, where the detector (tensor) response $D^{ij}$ depends on the detector's design and characteristics. Introducing the pattern functions $F_A(\nvh_\alpha)=D^{ij}e_{ij}^A(\nvh_\alpha)$, we can write the continuous (scalar) output $h(t)$ of the detector as
\be
    h(t) = \int_{-\infty}^{+\infty} \!\dd f \sum_{\alpha=1}^{N_0}\,\tilde h_\alpha(f,\nvh_\alpha) \,\mathrm{e}^{-2\pi \mathrm{i} f t}
\ee
where $\tilde h_\alpha(f,\nvh_\alpha)=\sum_A F_A(\nvh_\alpha) \tilde h_{\alpha A}(f,\nvh_\alpha)$. Since GW detectors have (very) limited angular resolution, we will be primarily interested in the statistics of
\be
    \label{eq:tildehf}
    \tilde h(f)\equiv \sum_{\alpha=1}^{N_0} h_\alpha(f,\nvh_\alpha) \;.
\ee
For a large number $N_0\gg 1$ of (mostly) weak, independent and unresolved sources, this GW strain is stochastic \citep[see the discussion in][]{romano/cornish:2017} and the Fourier modes $\tilde h(f)$ are random variables characterized by their statistical correlators.

Since the continuous detector output $h(t)$ is (uniformly) sampled at discrete times $t_0\leq t_n\leq t_0+T$, $n=0,1,\dots,N-1$, we introduce the discrete Fourier transform (DFT)
of the time domain signal computed at discrete frequencies $n \Delta f$, where $\Delta f=1/T$ is the fundamental frequency and $0\leq n<N$ is an integer such that $T/N=\Delta t$ is the sampling time.\footnote{The maximum measurable frequency is the Nyquist frequency $f_\text{Ny}=1/2\Delta t$.}
In the limit $T\to\infty$, all the Fourier modes are sampled and the discrete summations can be replaced by integrals. For our Fourier convention, the correspondence is $T\delta^\mathrm{K}_{f,f'}\to (2\pi)\delta^\mathrm{D}(f-f')$ and $\frac{1}{T}\sum_f \to \frac{1}{2\pi}\int\!\dd f$, where $\delta^\mathrm{K}$ and $\delta^\mathrm{D}$ are the Kronecker symbol and the Dirac distribution, respectively. With this definition, we choose
\be
    \label{eq:DFTdef}
    \tilde h_f  = \frac{\tilde h(f) * w_T(f)}{\sqrt{T}}\equiv \frac{\tilde h_T(f)}{\sqrt{T}}\;,
\ee
setting the convention for the DFT and its inverse. $\tilde h_f$ is the DFT measured from the discrete time series while $\tilde h_T(f)$ is the convolution of $\tilde h(f)$ with the spectral response $w_T(f)$ of the window function $w_T(r)$ (we will assume a simple rectangular window of length $T$ throughout). Note that $\tilde h_f$ has units of $\Hz^{-1/2}$ whereas $\tilde h(f)$ has units of $\Hz^{-1}$.

The power spectral density (PSD) $\hat P_h$ inferred from the discretized GW signal is computed through a suitable average over frequency bins, i.e.
\be
    \hat P_h(f) = \frac{1}{N(f_1,f_2)}\sum_{f_1\leq f\leq f_2}\big\lvert\tilde h_f\big\lvert^2\;,
\ee
which does not require a dimensional pre-factor due to our definition of the DFT.
Here $N(f_1,f_2)$ is the number of modes in the frequency range $f_1\leq f\leq f_2$.

$\hat P_h(f)$ provides an unbiased estimator for the actual (single-sided) PSD $S_h(f)$ of the GW signal,
\be
    \label{eq:psd}
    \big\langle \hat P_h(f)\big\rangle
    = \frac{F}{2}\, S_h(f)\;.
\ee
The brackets $\big\langle \cdot\big\rangle$ denotes an average over random realizations of the observed GW strain, $F=\big\langle F_+^2\big\rangle_{\nvh} +\big\langle F_\times^2\big\rangle_{\nvh}$ is the angular efficiency factor of the detector ($F=2/5$ for interferometers) and the factor $1/2$ guarantees that $\big\langle h^2(t)\big\rangle = \int_0^\infty \!\dd f\,S_h(f)$. For ergodic signals (which is the assumption we will make here since the statistical properties of $h(t)$ are stationary across realistic observational periods), the ensemble average can be estimated through a time average of the data. Note also that the units of $\hat P_h(f)$ and $S_h(f)$ are $\Hz^{-1}$.

For Gaussian fluctuations, the quantity $S_h(f)$ completely specifies the statistical properties of the measured Fourier modes $\tilde h_f$.

\subsection{Characteristic function}
\label{sec:characteristicfunction}

We follow \cite{Ginatetal2020} and derive the distribution $P(\tilde h_f)$ at observed frequencies $f\geq 0$ from the characteristic function. Since $\tilde h_f$ is a complex variable, the single source characteristic function is the expectation value
\be
    \label{eq:psi1}
    \psi_f(\tilde q) = \mathbb{E}\big[\mathrm{e}^{\mathrm{i}\,\Re(\tilde q^* \tilde h_{f,\nvh})}\big]\;,
\ee
where $\Re(\cdot)$ designate the real part of a complex number, $\tilde q\in\mathbb{C}$,
\be
    \label{eq:singlesourceDFT}
    \tilde h_{f,\nvh}\equiv \frac{\tilde h_T(f,\nvh)}{\sqrt{T}}
\ee
is the DFT of the single source GW signal, and the ensemble average $\mathbb{E}[~\cdot~]$ is taken over the source parameter space and, thereby, depends on the nature of the sources.

For GW signals produced by compact binary mergers for instance, the single source Fourier amplitude $\tilde h_{f,\nvh} = \tilde h_{f,\nvh}(t_0,T,\vxi,\vr,\varpi)$ is a function of $(t_0,T)$ as well as the intrinsic source parameters denoted by the vector $\vxi$, which includes the binary formation time $t_*$, the initial period $T_*$, the chirp mass $M_c$ etc. In addition, it depends on the three-dimensional co-moving position $\vr=(r,\nvh)$ of the source (on the past-light cone of the observer), and on the orbital phase $\varpi$. Since the latter is uniformly distributed in the range $0\leq\varpi<2\pi$, the ensemble average Eq.~(\ref{eq:psi1}) thus reads
\be
    \label{eq:psi2}
    \psi_f(\tilde q) =
    \int\! \dd^3\vr \int\! \dd\vxi\, \phi(\vr,\vxi) \int\!\frac{\dd\varpi}{2\pi}\, \mathrm{e}^{\mathrm{i}\,\Re(\tilde q^* \tilde h_{f,\nvh})}\;,
\ee
in which $\mathrm{d}^3\boldsymbol{r}=r^2\mathrm{d} r\; \mathrm{d}^2\hat{\boldsymbol{n}}$ is the infinitesimal co-moving volume, $\dd\vxi$ is the measure in the source parameter space, and $\phi(\vr,\vxi)$ is the joint PDF for the parameters $(\vr,\vxi)$. We will hereafter assume $0\leq r\leq r_0$ where the cutoff scale $r_0$ can be set to e.g. the radial co-moving radius of the Universe, $r_0=13.8\gpc$. Furthermore, although our approach can incorporate clustered sources, we will restrict ourselves to a spatial Poisson process\footnote{This is an excellent approximation when the distance between the source and observer is much larger than the characteristic clustering length.} and set $\phi(\vr,\vxi)\equiv\phi(\vxi)$ in practical computations.

The integral over $\varpi$ can be performed in Eq.~(\ref{eq:psi1}), because $\tilde{h}_{f,\nvh} \propto \mathrm{e}^{\mathrm{i}\varpi}$ and leads to
\begin{equation}
\begin{aligned}
    \psi_f(\tilde q)
    &= \int\!\dd^3\vr \int\!\dd\vxi\, \phi(\vr,\vxi)\, J_0(q|\tilde h_{f,\nvh}|) \\ &\equiv \psi_f(q)\;,
\end{aligned}
\end{equation}
which depends on the modulus $q=\abs{\tilde q}$ solely like the corresponding time domain characteristic function \cite{Ginatetal2020}. In other words, all the information about the phase of the GW signal is lost.

Assuming that the sources are identical and their total number in the Universe obeys a Poisson distribution of mean $N_0$, the characteristic function $\psi_f^{(N_0)}(q)$ of all the sources is a Poisson mixture of the single source $\psi_f(q)$. It can be recast into the form
\begin{align}
\label{eq:psiN0}
    \psi_f^{(N_0)}\!(q) &= \mathrm{e}^{-N_0}\sum_{n=0}^\infty \frac{N_0^n}{n!} \psi_f(q)^n \\
    &= \mathrm{e}^{N_0 G(q;f)} \nonumber\;,
\end{align}
with the generating function
\be
    \label{eqn:G definition}
    G(q;f)=\int\dd^3\vr \int\!\dd\vxi\, \phi(\vr,\vxi)\, \Big(J_0\big(q|\tilde h_{f,\nvh}|\big)-1\Big)\;.
\ee
The function $G(q;f)$ generically is a negative, monotonically decreasing function of $q\geq 0$, with $G(0;f)=0$. In the limit $q\to\infty$, $G(q;f)$ does not asymptote to $-1$ because the DFT $\tilde h_{f,\nvh}$ can vanish for a (significant) fraction of the parameter space (cf. \S\ref{sec:DFT}). As a result, $G(q;f)$ asymptotes to a (negative) constant which depends on the duration $T$ of the experiment (cf. \S\ref{sec:toymodel}).

\subsection{1-point distribution function}
\label{sec:PDF}

The (1-point) probability distribution function (PDF) of the observed DFT follows by Fourier transformation of $\psi_f^{(N_0)}(q)$,
\be
    \label{eq:PDFh1}
    P\big(\tilde h_f\big) = \frac{1}{(2\pi)^2}\int\!\dd^2\tilde q\,
    \mathrm{e}^{-\mathrm{i}\Re(\tilde q^*\tilde h_f)}\,\psi_f^{(N_0)}(q) \;.
\ee
For a stationary GW signal, higher-point distribution functions contain no further information.
Because $G(q;f)$ tends toward a constant $G_\infty(f)<0$ in the limit $q\to\infty$, this integral is formally divergent. In the following, we demonstrate the regularization procedure and identify the physical interpretation of the different terms.

\subsubsection{Extracting the finite part of $P(\tilde h_f)$}

To extract the finite-part of Eq.~(\ref{eq:PDFh1}), observe that, for each possible number $n$ of sources, the probability of finding a gravitational-wave strain $\tilde h_f$ is
\begin{equation}
    P_n\big(\tilde h_f\big) = \frac{1}{(2\pi)^2}\int\!\dd^2\tilde q\,\mathrm{e}^{-\mathrm{i}\Re(\tilde{q}^*\tilde{h}_f)}\,\psi_f(q)^n\;,
\end{equation}
while the full probability distribution is
\begin{equation}
    P\big(\tilde h_f\big) = \mathrm{e}^{-N_0}\,\sum_{n=0}^{\infty}\frac{N_0^n}{n!}\,P_n\big(\tilde h_f\big) \;.
\end{equation}
The first two terms in the sum are the probabilities for detecting a gravitational wave $\tilde h_f$ given that there are no sources or exactly one source, respectively.

The first term ($n=0$) reads
\begin{equation}
    \mathrm{e}^{-N_0}\, P_0\big(\tilde h_f\big)
    = \mathrm{e}^{-N_0}\,\delta^\text{D}\!\big(\tilde h_f\big)
    = \frac{\mathrm{e}^{-N_0}}{2\pi h}\,\delta^\text{D}(h)\;.
\end{equation}
This is easy to interpret physically: if there are no sources, the gravitational wave amplitude must be zero, deterministically. The second term ($n=1)$ is
\begin{equation}
    \label{eq:P1_0}
    P_1\big(\tilde{h}_f\big) = \frac{1}{(2\pi)^2}\int\!\dd^2 \tilde q\, \mathrm{e}^{-\mathrm{i}\Re(\tilde{q}^*\tilde{h}_f)}\,\psi_f(q) \;.
\end{equation}
Eq.~(\ref{eq:psi2}) shows that strictly speaking, this expression does not converge in the sense of functions, but it does converge distributionally. To see this, let $\tilde h_f = h \mathrm{e}^{i\theta}$ and $\tilde q = q \mathrm{e}^{i\alpha}$ be the polar form of the complex variables $\tilde h_f$ and $\tilde q$, respectively (we will use the notation $h\equiv |\tilde h_f|$ throughout whenever it is not confusing).
Upon integrating out the phase $\alpha$, Eq.~(\ref{eq:P1_0}) becomes
\begin{align}
    P_1\big(\tilde h_f\big) &= \frac{1}{(2\pi)^2}\int_0^\infty\!\dd q\,q\int_0^{2\pi}\!\dd\alpha\,\mathrm{e}^{-\mathrm{i}\Re(\tilde{q}^*\tilde{h}_f)}\,\psi_f(q) \nonumber \\
    &= \frac{1}{2\pi}\int_0^\infty\!\dd q\,q\,J_0(qh)\,\psi_f(q)
    \nonumber \\
    &= \frac{1}{2\pi}\mathbb{E}\left[\int_0^\infty\!\dd q\, q \,J_0(qh)\,J_0\big(q|\tilde h_{f,\nvh}|\big)\right]
    \nonumber \\
    &= \frac{1}{2\pi h}\mathbb{E}\Big[\delta^\text{D}\!\big(h-|\tilde h_{f,\nvh}|\big)\Big]
    \label{eq:P1_1} \;.
\end{align}
Here, we have used the generalized integral
\begin{equation}
    \int_0^\infty\!\dd x\, x\, J_0(xa)\, J_0(xb)= \frac{\delta^\text{D}(a-b)}{a}\;,
\end{equation}
which can be derived from equation (10.22.62) of \cite{DLMF} by taking the limit $\mu \to \nu = 0$. Here, the single-source expectation value $\mathbb{E}[~\cdot~]$ can be taken over the parameters $(\vr,\vxi)$ solely since  $|\tilde h_{f,\nvh}|=\tilde h_{f,\nvh}(t_0,T,\vr,\vxi)$ is independent of the orbital phase $\varpi$.

While the $n=0$ contribution is anomalous already in the time-domain analysis \cite{Ginatetal2020}, it can be safely ignored here, too, as it only contributes at $h=0$. The anomalous, $n=1$ case is new and arises because $\tilde{h}_f$ is a complex number, whereas $h(t)$ is real. \begin{widetext}
With these expressions, we can write
\begin{align}
    \label{eqn:probability of h tilde renormalised}
    P\big(\tilde h_f\big) &= \frac{\mathrm{e}^{-N_0}}{2\pi h}\Big\{\delta^\text{D}(h) + N_0\,\mathbb{E}\Big[\delta^\text{D}\big(h- |\tilde h_{f,\nvh}|\big)\Big]\Big\}
    + \frac{1}{(2\pi)^2}\int\!\dd^2\tilde q \, \mathrm{e}^{-\mathrm{i}\Re\left(\tilde{q}^*\tilde{h}_f\right)}\left(\sum_{n=2}^{\infty}\frac{\mathrm{e}^{-N_0}N_0^n}{n!} \psi_f(q)^n\right) \\
    &= \frac{\mathrm{e}^{-N_0}}{2\pi h}\Big\{\delta^\text{D}(h) + N_0\,\mathbb{E}\Big[\delta^\text{D}\big(h- |\tilde h_{f,\nvh}|\big)\Big]\Big\}
    + \frac{1}{(2\pi)^2}\int\!\dd^2\tilde q\, \mathrm{e}^{-\mathrm{i}\Re\left(\tilde{q}^*\tilde{h}_f\right)}\left[\mathrm{e}^{N_0G(q;f)} - \mathrm{e}^{-N_0}\big(1+N_0 + N_0G(q;f)\big)\right]\;, \nonumber
\end{align}
where, in the second equality, we have carried out the sum over $n$.

\subsubsection{Fourier modes and phase distributions}

Upon writing $\dd^2\tilde h_f =h \dd h \dd\theta$, the fact that $G$ is a function of $q$ only implies that $P(\tilde h_f)\,\dd^2\tilde h_f=P(h;f)\,P(\theta;f)\,\dd h \dd\theta$, with
\begin{equation}
    \label{eqn:P(|h|) bessel}
    P(h;f) = \mathrm{e}^{-N_0}\Big\{\delta^\text{D}(h) + N_0\,\mathbb{E}\Big[\delta^\text{D}\big(h- |\tilde h_{f,\nvh}|\big)\Big]
    \Big\} + h\int_0^{\infty}\!\dd q\, q J_0(qh)\left[\mathrm{e}^{N_0G(q;f)} - \mathrm{e}^{-N_0}\big(1+N_0 + N_0G(q;f)\big)\right]
\end{equation}
\end{widetext}
which follows from \eqref{eqn:probability of h tilde renormalised}, and
\be
    P(\theta;f) = \frac{1}{2\pi} \;.
\ee
As expected, $\theta$ is uniformly distributed in the range $0\leq\theta < 2\pi$.

One may want to decompose $P(h;f)$ as
\begin{equation}
    P(h;f) \equiv \mathrm{e}^{-N_0}P_0(h;f) + N_0\mathrm{e}^{-N_0}P_1(h;f) + P_\text{many}(h;f)
\end{equation}
where $P_\text{many}(h;f)$ denotes the integral in the right-hand side of Eq.~(\ref{eqn:P(|h|) bessel}).
Note that by equation \eqref{eqn:G definition}, the terms in the square brackets in $P_\text{many}$ behave like $O(q^{-1})$ as $q \to \infty$, for fixed $N_0$, and therefore the integral converges. For $N_0\gg 1$ however, the $n=0$ and 1 terms are exponentially suppressed, so that separating the sum into different pieces is not needed for $h>0$; in this regime, $P(h;f) \approx P_\text{many}(h;f)$ is simply given by
\be
P(h;f) = h\int_0^\infty\! \mathrm{d}q\,q\,J_0(qh)\,\mathrm{e}^{N_0 G(q;f)} \;,
\ee
which is the Hankel transform (of order 0) of the function $\mathrm{e}^{N_0G(q;f)}$.

As a sanity check, on replacing $\mathrm{e}^{N_0G(q;f)}$ by the characteristic function $\mathrm{e}^{-\frac{1}{8} q^2 S_h(f)}$ of a Gaussian SGWB signal, the Hankel transform returns, as expected, the Rayleigh distribution
\be
    P(h;f) = \frac{4h}{S_h(f)}\, \mathrm{e}^{-2 h^2/S_h(f)}\;,
\ee
which has a second moment $\langle h^2\rangle = \frac{1}{2} S_h(f)$.

For numerical evaluation, we follow \cite{Ginatetal2020} and introduce the dimensionless variables $s=q h_c$, and $x=h/h_c$. Here, $h_c$ is a (possibly frequency-dependent) characteristic DFT amplitude and thus has unit of $\Hz^{-1/2}$. This allows us to write down the characteristic function $G$ and the PDF $P$ as
\begin{align}
    G(s;f) &= \int\!\dd^3\vr\int\!\dd\vxi\, \phi(\vr,\vxi) \left(J_0\left( s\frac{|\tilde h_{f,\nvh}|}{h_c}\right)-1\right) \nonumber \\
    P(x;f) &= x\int_0^\infty\!\dd s\, s\, J_0\big(s x\big)\, \mathrm{e}^{N_0 G(s;f)} \label{eq:sxvariables} \;,
\end{align}
where $P(h;f) = P(x;f)/h_c$.
In practice, it is convenient to choose $h_c\sim \big\langle h\big\rangle$.

Before concluding this section, we emphasize that $N_0$ counts all the sources that have formed by the retarded time $t_{0,\ret}=t(\eta_0-r/c)$ in a co-moving volume $V_0=(4\pi/3)r_0^3$ centered on the observer. Therefore, $N_0$ includes also sources that have already merged (for which $h(t)\equiv 0$).
When $N_0\gg 1$, as is the case of the GW signal produced by compact binary mergers in the Universe, $P(h;f)$ is very close to a Gaussian distribution, for long observation times, in accordance with the classical central limit theorem (CLT), \citep[e.g.,][]{Berry:1941,Esseen:1942,BORGERS2018679} which guarantees the point-wise convergence of a sum of identically distributed variables with finite variance, higher order moments are suppressed by powers of $N_0$ by means of an Edgeworth expansion, i.e. the generalization of the CLT.\footnote{A divergence would arise from the far-field source distribution (Olber's paradox) if it were infinite. It is absent for realistic SGWBs because there are no sources beyond the radius of the observable Universe. This far-field cutoff was not taken into account in the discussion of \cite{Ginatetal2020} about CLTs.} Notwithstanding, $P(h;f)$ converges non-uniformly towards a Gaussian owing to the emergence of a high-strain power-law tail, which is produced by bright, close sources as discussed in \S\ref{sec:asymptotics} below.

\subsection{Second moment and GW energy spectrum}
\label{sec:SGWBspectrum}

Moments of the observed DFT can be obtained by taking derivatives of the generating function $G(q;f)$.
In particular, the second moment $\langle |\tilde h_f|^2\rangle$ of the distribution $P(|\tilde h_f|)$ is given by
\begin{align}
    \langle |\tilde h_f|^2\rangle & = \int \dd^2\tilde{h}_f \, |\tilde{h}_f|^2\, P(\tilde{h}_f) \nonumber \\ &
    = \frac{1}{(2\pi)^2}\int\! \dd^2\tilde{h}_f\int\! \dd^2 \tilde q\, |\tilde{h}_f|^2 \mathrm{e}^{-\mathrm{i}\Re(\tilde q^*\tilde{h}_f)}\mathrm{e}^{N_0G(q;f)} \nonumber \\ &
    = -2\, N_0\, G''(0;f)
\end{align}
or, equivalently,
\begin{align}
    \label{eq:hf2average}
    \langle |\tilde h_f|^2\rangle &=N_0\,\mathbb{E}\big[|\tilde h_{f,\nvh}|^2\big] \\
    &=N_0 \int\! \dd^3\vr \int\! \dd\vxi\, \phi(\vr,\vxi)\, |\tilde h_{f,\nvh}|^2 \nonumber \;.
\end{align}
This result agrees with a derivation based on the distribution of the real and imaginary parts of $\tilde h_f$, which returns $\langle \Re(\tilde h_f)^2\rangle=\langle\Im(\tilde h_f)^2\rangle = -N_0 G''(0;f)$.\footnote{The characteristic function $\psi_f^{(N_0)}(q)$ for $\Re(\tilde h_f)$ (resp. $\Im(\tilde h_f)$) is identical to Eq.~(\ref{eq:psiN0}) except that $|\tilde h_f|$ is replaced by $\Re(\tilde h_f)$ (resp. $\Im(\tilde h_f)$) and the domain of the real variable $q$ is the whole real axis. The equality $\langle \Re(\tilde h_f)^2\rangle=\langle\Im(\tilde h_f)^2\rangle = -N_0 G''(0;f)$ follows from a simple one-dimensional Fourier transform.} The definition Eq.~(\ref{eq:psd}) of the single-sided PSD then implies
\be
    \label{eq:FSh}
    F S_h(f) = 2 N_0 \int\! \dd^3\vr \int\! \dd\vxi\, \phi(\vr,\vxi)\, |\tilde h_{f,\nvh}|^2\;,
\ee
which can be used to compute the (dimensionless) GW energy density
\be
\ogw(f) = \frac{1}{\rho_c}\frac{\dd\rho_\text{gw}}{\dd\ln f}
= \frac{4\pi^2}{3H_0^2}f^3 S_h(f)
\ee
for any type of discrete superposition of GW sources. Here, $\rho_c=3c^2 H_0^2/8\pi G$ is the present-day critical density.

Note that the PSD $S_h(f)$ is independent of $T$. We will illustrate this point explicitly in \S\ref{sec:toymodel}.

\subsection{Large-Strain Asymptotics}
\label{sec:asymptotics}

For large values of $h$, the distribution of observed Fourier modes is dominated by a small number of bright sources with little or no destructive interference. These sources can be subtracted from the signal (see \S\ref{sec:confusionnoise}).

Let us derive the large-$x$ behavior of $P(x;f)$ when these sources are present.
As in the time domain, \cite{Ginatetal2020}, the frequency space generating function \eqref{eq:sxvariables} can be expanded in the series
\be
    \label{eqn:G small s approximation}
    G(s;f) \sim -a(f)\,s^2 + b(f)\, |s|^3 - d(f)\, s^4 + \ldots \;,
\ee
in the neighborhood of $s=0$.
This small-$s$ expansion may be obtained by applying the same techniques as in Refs. \cite{Ginatetal2020,Konradetal2022} (see also references therein). See appendix \ref{app:G approximation} for a derivation of the expressions of $a$ and $b$. Inserting this approximation into the probability distribution \eqref{eq:sxvariables}, we find 
\begin{align}
    P(x;f) & = x \int_0^\infty\!\!\! \dd s\, s\,J_0(sx)\,\mathrm{e}^{N_0G(s;f)}  \\ &
    \sim x\int_0^\infty \!\!\!\dd s\,\left[sJ_0(sx)\mathrm{e}^{-N_0as^2}\sum_{n=0}^\infty \frac{(N_0 bs^3)^n}{n!}\right] + \ldots \nonumber \\ &
    \sim x\int_0^\infty \!\!\!\dd s\,\Big[sJ_0(sx)\mathrm{e}^{-N_0as^2}(1+bN_0s^3)\Big]+ \ldots \nonumber
\end{align}
In the last expression, the first term yields the Rayleigh distribution, which decays exponentially at large $x$;
for the second, define $y=sx$, whence
\begin{align}
    P(x;f) & \sim \frac{x\,\mathrm{e}^{-x^2/(4N_0a)}}{2N_0a} + \frac{N_0b}{x^4}\int_0^\infty\!\dd y\, y^4 J_0(y)\,\mathrm{e}^{-\frac{N_0 a y^2}{x^2}} \nonumber \\ &
    = \frac{x\,\mathrm{e}^{-x^2/(4N_0a)}}{2N_0a} \label{eq:asymptotic} \\ &
    \qquad + \frac{N_0b}{x^4}\left(\frac{\partial^2}{\partial t^2} \int_0^\infty\!\dd y\, J_0(y)\, \mathrm{e}^{-ty^2}\right)_{t=N_0a/x^2}  \nonumber \;.
\end{align}
This integral may be evaluated \cite{DLMF}, to yield
\begin{align}
    \label{eqn: P_f asymptotic with Bessels}
    P(x;f) &\sim \frac{x\,\mathrm{e}^{-x^2/(4N_0a)}}{2N_0a} + \frac{\sqrt{\pi } b h^5 N_0 \mathrm{e}^{-\frac{h^2}{8a N_0}}}{64 (a N_0)^{9/2}} \\
    &\quad \times \bigg[\left(\frac{12 a N_0}{h^2} \left(\frac{2 a N_0}{h^2}-1\right)+1\right) I_0\left(\frac{h^2}{8 N_0 a}\right) \nonumber \\
    &\qquad +\left(\frac{8 a N_0}{h^2}-1\right) I_1\left(\frac{h^2}{8 N_0 a}\right)\bigg] \nonumber \;,
\end{align}
where $I_\nu(z)$ is the modified Bessel function of the first kind. The asymptotics of $I_\nu(z)$ at $z\to \infty$ are given by \cite{DLMF}
\be
    \mathrm{e}^{-z}\, I_0(z) \sim \frac{1}{\sqrt{2\pi z}}\sum_{k=0}^\infty \frac{a_k(\nu)}{z^k} \;,
\ee
where $a_0(\nu) = 1$ and
\be\label{eqn:a k of nu}
    a_k(\nu) = \frac{\prod_{n=1}^k[4\nu^2-(2k-1)^2]}{k!8^k} \;.
\ee
One can insert this expansion into equation \eqref{eqn: P_f asymptotic with Bessels}, and expand at large $h$, to find
\be
    \label{eq:highstrainlaw}
    P(x;f) \sim \frac{9 N_0 b(f)}{x^4}\;,
\ee
as $x\to \infty$.
Here, $b(f)$ is again the coefficient of
$|s|^3$ in the small-$s$ expansion of $G(s;f)$. The $h^{-4}$ power-law behavior discovered by \citet{Ginatetal2020} for the time domain probability distribution remains true in frequency space. This power law is universal, and the type/ model of astrophysical GW signal studied only affects the coefficient of $h^{-4}$, not the power law. The latter is in fact a consequence of the $1/$distance law of propagation of GWs in general relativity.

\section{Application to compact mergers}
\label{sec:compact mergers}

In this Section, we demonstrate the applicability of our approach with the GW signal produced by compact binary coalescences in the Universe. We assume an FLRW background where the source and observer are co-moving and thus share the same cosmic time $t$. We set the present-day scale factor $a(t_0)$ to unity. It will be convenient to work with the conformal time $\eta=\int\!\dd t/a(t)$, such that the experiment is carried out in the conformal time interval $\eta_0\leq\eta\lesssim \eta_0+T$ in the detector's frame (assuming $T \ll H_0^{-1}$). Furthermore, we will assume throughout a rectangular window $w_T(t)\equiv \Pi_T(t)$ (as an approximation to more realistic windows) for simplicity.

\subsection{In-spiral of a compact binary}
\label{sec:compactbinaries}

Our fiducial model approximates the sources as circular binaries -- an assumption justified by the circularizing effect of gravitational-wave emission. Temporarily ignoring the finite observation time $T$, the detector measures a sky-averaged, linear superposition $\tilde{h}(f,\nvh) = F_+(\nvh)\tilde{h}_+(f,\nvh) + F_{\times}(\nvh)\tilde{h}_{\times}(f,\nvh)$ of the two polarizations. For a single source, we have
\be
    \label{eq:fourierhfn}
    \tilde h(f,\nvh) = h_0(f)\, \mathrm{e}^{\mathrm{i}\Psi_+(f)}\, Q(\vartheta,\varphi,\imath)\;,
\ee
where, here and henceforth, we omit the dependence of $h_0$ on $(r,\vxi)$ to avoid clutter.
Here, $0\leq\imath\leq\pi$ is the inclination of the binary orbit relative to the line of sight direction $\nvh=(\vartheta,\varphi)$.
The amplitude $h_0(f)$ and phase $\Psi_+(f)$ are given by~\footnote{In ref. \cite{Maggiore}, $\Psi_+$ contains an additional $-\Phi_0$ term, i.e. minus the phase at coalescence, but here, as we have integrated over the global phase $\varpi$, this quantity is removed.}
\begin{align}
  & h_0(f) = \frac{1}{\pi^{2/3}}\sqrt{\frac{5}{24}}\frac{c}{d_L(z)}\left(\frac{G(1+z)M_c}{c^3} \right)^{5/6} f^{-7/6}\label{eq:gwamplitude} \\ &
  \Psi_+(f) = 2\pi f\left(t_{\textrm{coal}} + \frac{r}{c}\right) -\frac{\pi}{4} + \frac{3}{4}\left(\frac{8\pi GM_c f}{c^3}\right)^{-5/3} \;.
  \label{eq:gwphase}
\end{align}
The strain amplitude decreases with increasing luminosity distance $d_L(z)=(1+z) r$. Here, $z=z(r)$ describes the source redshift, $t_\text{coal}$ is the time at which the coalescence is detected by the observer (through the arrival of the GW signal), and
\be
    M_c = \frac{(m_1m_2)^{3/5}}{(m_1+m_2)^{1/5}}
\ee
is the chirp mass. Finally, the form factor,
\be
    \label{eq:Q}
    Q(\vartheta,\varphi,\imath) = F_+(\vartheta,\varphi)\frac{1+\cos^2\imath}{2} + \mathrm{i}\, F_{\times}(\vartheta,\varphi)\cos\imath\;,
\ee
encodes the dependence of the measured Fourier modes on the binary orientation and sky direction. For an interferometer with arms along the $x$ and $y$ axes, the form factors are given by $F_+(\vartheta,\varphi)=(1/2)(1+\cos^2\vartheta)\cos 2\varphi$ and $F_{\times}(\vartheta,\varphi)=\cos\vartheta\sin 2\varphi$ (with the convention that positive $+$ polarization is along the $\xhat$ axis).

The frequency $f_o(\eta,r,\vxi)$ of the GW signal detected at conformal time $\eta$,
\be
    \label{eq:obsgwfrequency}
    f_o(\eta,r,\vxi) = \frac{1}{(1+z)\pi}\left(\frac{5}{256\,\mathfrak{t}_s}\right)^{3/8}\left(\frac{GM_c}{c^3}\right)^{-5/8}
\ee
grows monotonically with time until the coalescence phase. It depends on $z$, $M_c$ and the time to coalescence $\mathfrak{t}_s$ as measured with the source's clock~\footnote{When the time to coalescence is much smaller than the Hubble time, i.e. $\mathfrak{t}_s\ll H^{-1}$ as is the case of binaries about to merge, one can set $(1+z)\mathfrak{t}_s\approx \mathfrak{t}_o$ in Eq.~(\ref{eq:obsgwfrequency}) where $\mathfrak{t}_o$ is the time to coalescence measured in the detector's frame.}
\be
\label{eq:timetocoalescence}
\mathfrak{t}_s(\eta,r,\vxi) = t_*+\tau_0(M_c,T_*) - t(\eta - r/c) \;.
\ee
The total lifetime of the source is the time $\tau_0$ to coalescence at formation, which is given by (e.g. \citep{Maggiore})
\begin{align}
    \tau_0(M_c,T_*) &= \frac{5}{256}\,c^5\, \left(\frac{T_*}{2\pi}\right)^{8/3}\, \big(G M_c\big)^{-5/3} \\
    &\simeq  3.226\times 10^{17}\yr\left(\frac{T_*}{\yr}\right)^{8/3}\left(\frac{\msun}{M_c}\right)^{5/3}
    \nonumber \;.
\end{align}
This is the lifetime of a binary with an initial period $T_*$.

The wave-form Eqs.~(\ref{eq:gwamplitude}) -- (\ref{eq:gwphase}) (produced by a slow adiabatic sequence of circular orbits) is only valid for frequencies $f_o(\eta,r,\vxi)\lesssim f_\text{merg}(r,\vxi)\sim \fisco(r,\vxi)$~\footnote{$\fisco(r,\vxi)= 2200\Hz\times \frac{M_\odot}{m_1+m_2}$ is the observed GW frequency corresponding to the innermost stable circular orbit or ISCO.} above which strong-field GR effects cannot be neglected. We use the following template to extend the GW signal through the merger and ring-down phase (and avoid truncating the signal at ISCO) \cite[see, e.g.,][]{ajith/etal:2008,marassi/schneider/etal:2011}:
\be
    \tilde h(f) = h_\text{eff}(f)\, \mathrm{e}^{\mathrm{i}\Psi_\text{eff}(f)}\, Q(\vartheta,\varphi,\imath)\;.
\ee
The effective Fourier amplitude is given by $h_\text{eff}(f) = h_0(f_\text{merg}) A_{\rm GR}$, with
\begin{align}
    \label{eq:FTamplitudes}
    A_{\rm GR} \equiv \left\lbrace
    \begin{array}{cc} \bigg. \left(\frac{f}{f_\text{merg}}\right)^{-7/6} & f < f_\text{merg} \\
    \left(\frac{f}{f_\text{merg}}\right)^{-2/3} & f_\text{merg} < f < f_\text{ring} \\ \bigg.
    \omega\mathcal{L} & f_\text{ring} < f < f_\text{cut}\;,
    \end{array}\right.
\end{align}
where
\begin{align}
    \omega &= \frac{\pi\sigma}{2}\left(\frac{f_\text{ring}}{f_\text{merg}}\right)^{-2/3} \\
    \mathcal{L} &= \frac{1}{2\pi}\frac{\sigma}{(f-f_\text{ring})^2+\sigma^2/4} \nonumber \;.
\end{align}
The frequencies $f_\text{merg}(r,\vxi)$, $f_\text{ring}(r,\vxi)$, $f_\text{cut}(r,\vxi)$ and $\sigma(r,\vxi)$ defined in the detector's rest frame are functions of the binary component masses $m_1$ and $m_2$ as given in \cite{ajith/etal:2008}.
The explicit expression for $\Psi_\text{eff}(f)$ is not needed here since $P(\tilde h_f)$ is independent of the phase.

\subsection{DFT and the stationary point condition}
\label{sec:DFT}

For a finite observation time $T$ (as measured in the detector's rest frame), the DFT $\tilde h_{f,\nvh}$ of the single source GW signal can be calculated with the stationary-phase method \cite[e.g.][]{Maggiore}, as a time integral. Then, $\tilde h_{f,\nvh}$ is non-vanishing only if the stationary point $\bar t$ falls inside the observation window $[t_0,t_0+T]$. Translating into conformal time, we want $\bar\eta\equiv \eta(\bar t)$ to satisfy
\be
    \label{eq:stationarycondition}
    \eta_0\leq \bar\eta\leq \eta_0+T \;.
\ee
The stationary point associates the observed frequency $f$ with the GW source parameters via Eq.~\eqref{eq:obsgwfrequency}: $f\equiv f_o(\mathfrak{t}_s,r,\vxi)$. This relation can be inverted to find $\mathfrak{t}_s(f,r,\vxi)$, which is given by
\begin{equation}
    \mathfrak{t}_s(f,r,\vxi) \equiv \frac{5}{256}\Big({\pi f(1+z)\Big)^{-8/3}}\left(\frac{GM_c}{c^3}\right)^{-5/3}
\end{equation}
Substituting into Eq.~(\ref{eq:timetocoalescence}), we obtain $\mathfrak{t}_s(f,r,\vxi)=t_* + \tau_0 - t(\bar\eta-r/c)$ or, equivalently, $\bar\eta-r/c=\eta(\tau_0-\mathfrak{t}_s(f,r,\vxi)+t_*)$. Therefore, the condition Eq.~(\ref{eq:stationarycondition}) on the stationary point may also be stated as
\begin{equation}\label{ineq:heaviside condition}
    \eta_0 -\frac{r}{c} \leq \eta\left(\tau_0 - \mathfrak{t}_s(f,r,\vxi) + t_*\right) \leq \eta_0 -\frac{r}{c}+ T \;.
\end{equation}
To relate $\tilde h_{f,\nvh}$ to $\tilde h(f,\nvh)$ and the window function $w_T(t)$ of the experiment however, we need to rephrase the stationary point condition in terms of frequencies. This is straightforward:
\be
    f_o(\eta_0,r,\vxi)\leq f\leq f_o(\eta_0+T,r,\vxi) \;.
\ee
The DFT of the single source GW signal defined in Eq.~(\ref{eq:singlesourceDFT}) thus is
\be
    \label{eq:singlesourceDFT1}
    \tilde h_{f,\nvh} = \frac{\tilde h_T(f,\nvh)}{\sqrt{T}}\approx \tilde h(f,\nvh)\frac{\Pi_T(f)}{\sqrt{T}} \;,
\ee
where
\begin{align}
    \Pi_T(f)&= \Theta\big(f-f_o(\eta_0,r,\vxi)\big) \\
    &\qquad\times \Theta\big(f_o(\eta_0+T,r,\vxi)-f\big)
    \nonumber
\end{align}
is a rectangular function which approximates the window function of the experiment.
Note that $\Pi_T(f)^k = \Pi_T(f)$ for any integer $k\geq 1$. This can be used to simplify the expression of $G(q;f)$, see Eq.~(\ref{eq:ToyGsf}) below for instance.

\subsection{Merger rate and source counts}
\label{sec:rates}

The number density $\dd N/\dd f$ of overlapping sources (at the GW detector) per frequency is the expectation value
\begin{align}
    \frac{\dd N}{\dd f} &= N_0\, \mathbb{E}\Big[\delta^\text{D}\!\big(f-f_o(\eta_0,r,\vxi)\big)\Big] \\
    &= N_0 \int\!\dd^3\vr\int\!\dd\vxi\,\phi(\vr,\vxi)\,\delta^\text{D}\!\big(f-f(\eta_0,r,\vxi)\big) \nonumber
\end{align}
and has units of $\Hz^{-1}$. For the adiabatic sequences of circular orbits considered here, this differential number density scales as $\dd N/\dd f\propto f^{-11/3}$ and accounts for the time binaries spend in a given frequency bin \cite{finn/thorne:2000}.

Furthermore, the value of $N_0$ is constrained by the observed, present-day merger rate $R_\text{merger}$ of compact binaries per co-moving volume as inferred by GW experiments. Recent analyses of the Advanced {\small LIGO} and {\small VIRGO} third observing run (O3) yields $R_\text{merger}\sim 1000 \gpc^{-3}\yr^{-1}$ for binary BH and neutron star (NS) mergers \cite{LIGOvirgo2019PRX,LIGOvirgo2022}. In our approach, this merger rate can be derived from the requirement $\mathfrak{t}_s\equiv 0$. Concretely,
\be
    \label{eq:mergerrate}
    R_\text{merger} = \left(\frac{N_0}{V_0}\right) \int\!\dd^3\vr\int\!\dd\vxi\,\phi(\vr,\vxi)\,\delta^\text{D}\!\big(t_*+\tau_0-t_{0,\ret}\big)\;,
\ee
which has units of $\gpc^{-3}\yr^{-1}$.
Recall that $V_0=(4\pi/3)r_0^3$ is the co-moving volume of the present-day, observable Universe and $t_{0,\ret}=t(\eta_0-r/c)$ is the retarded time at emission. We use this equation to determine $N_0$.

\begin{figure}
    \centering
    \includegraphics[width=0.45\textwidth]{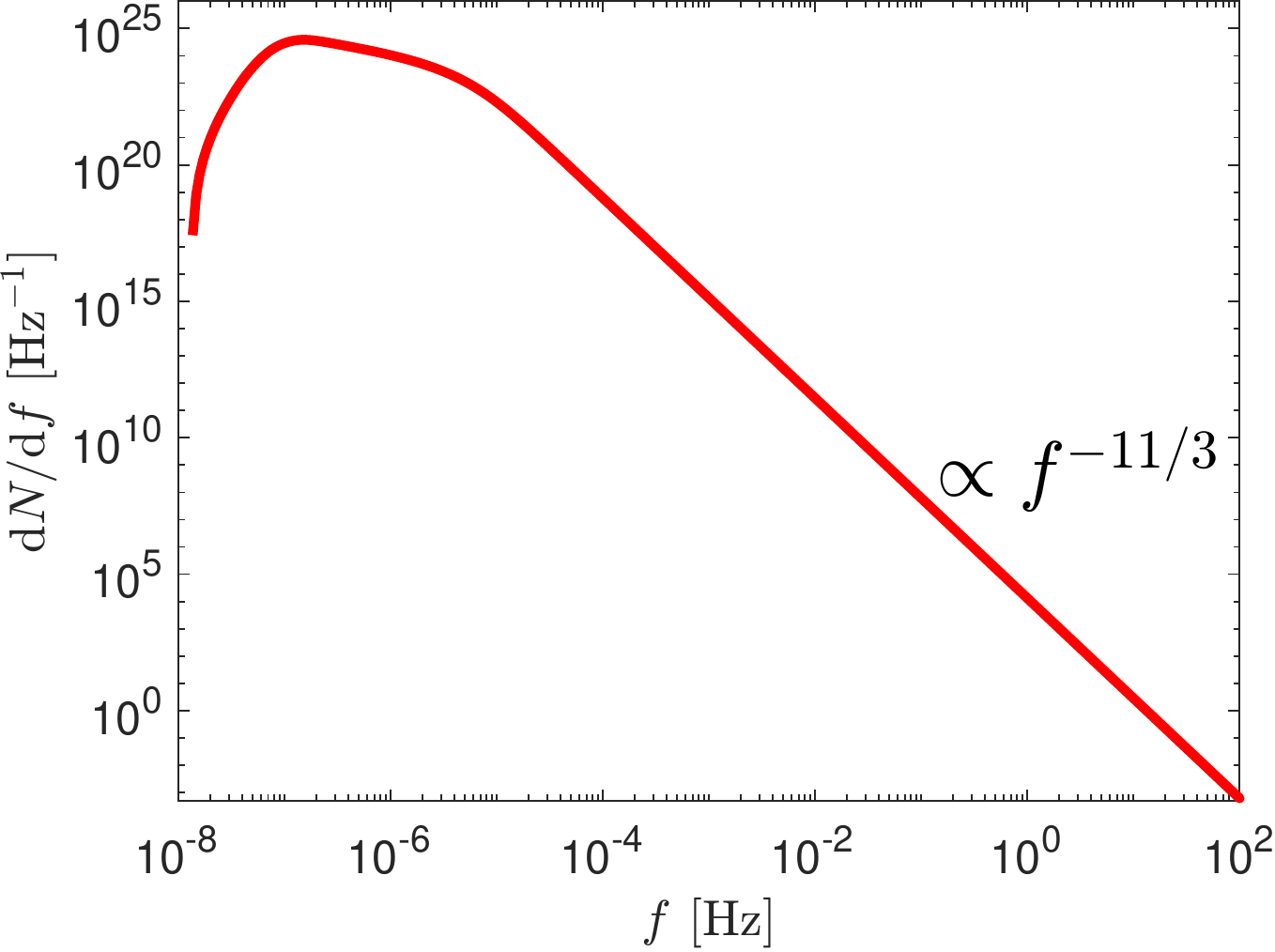}
    \caption{The differential number of sources per unit frequency, as given in equation \eqref{eqn:dNdf} in a toy model in which only the binary formation time $t_*$ and chirp mass $M_c$ are varied (see text).}
    \label{fig:dndf}
\end{figure}

\section{The Distribution of The SGWB}
\label{sec:results for probabilities}

Having specified the wave-form of the sources considered here, we are now at a position to apply the formalism spelt out in \S\ref{sec:theory} to compact binaries in the Universe. We start with a toy model to illustrate the main points (\S\ref{sec:toymodel}), and then move on to describe a more realistic model of the unsubtracted signal produced by all the sources in the Universe (\S\ref{sec:realisticSGWB}). The latter provides our basis for studying the resulting SGWB obtained by subtracting bright sources above the detection threshold (\S\ref{sec:confusionnoise}). 

\subsection{Insights from a simplified model}
\label{sec:toymodel}

To illustrate our approach, consider first a simplified model in which the source parameter vector $\vxi = (t_*,M_c,\imath)$ is limited to the binary formation time $t_*$, the chirp mass $M_c$ and the inclination $\imath$.
This allows us to write down relatively simple expressions for the different model ingredients. We begin by evaluating $N_0$, then we derive the generating function $G$, use it to calculate $P(x;f)$ and relate it to $\Omega_{\rm gw}$.

\subsubsection{Source counts}

For a spatial Poisson process, the joint PDF $\phi(\vr,t_*,M_c,i)$ reduces to
\be
    \phi(\vr,t_*,M_c,\imath) = \frac{3}{8\pi r_0^3}\,\phi(t_*)\,\phi(M_c)\;,
\ee
where the factor of $8\pi$ arises because the angular is $\dd^2\nvh\,\dd\cos\imath=\dd\cos\vartheta\,\dd\varphi\,\dd\cos\imath$. The distributions $\phi(t_*)$ and $\phi(M_c)$ are detailed in Appendix \ref{app:galaxymodel}. In short, $m_1, m_2\in [5,80] \msun$ with a power-law distribution of slope $-2.7$. Furthermore, we assume a single initial binary separation $a_*=0.01 \AU$, for the simplified model in this sub-section.

The total number of sources $N_0$ is constrained by the observed present-day merger rate Eq.~(\ref{eq:mergerrate}), which is
\begin{align}
    R_\text{merger}= & \;\left(\frac{N_0}{V_0}\right) \int_0^{r_0}\!\dd r\,r^2\int\!\dd^2\nvh\int\!\dd\cos\imath\,\int\!\dd M_c\int\!\dd t_* \nonumber \\
    & \times \frac{3}{8\pi r_0^3}\,\phi(t_*)\,\phi(M_c)\, \delta^D(t_*+\tau_0-t_{0,\ret}) \nonumber \\
    = &\; \frac{4\pi N_0}{V_0^2}\int_0^{r_0}\!\dd r\,r^2 \int\!\dd M_c \nonumber \\
    & \times\phi(t_*=t_{0,\ret}-\tau_0)\,\phi(M_c) \;.
\end{align}
Equating $R_\text{merger}$ to the merger rate inferred from the resolved mergers yields the model-dependent normalization $N_0\simeq 4.15\times 10^{18}$,
which is reasonable given the presence of $\mathcal{O}(10^{12})$ galaxies in our observable Universe.

\begin{figure}
    \centering
    \includegraphics[width=.45\textwidth]{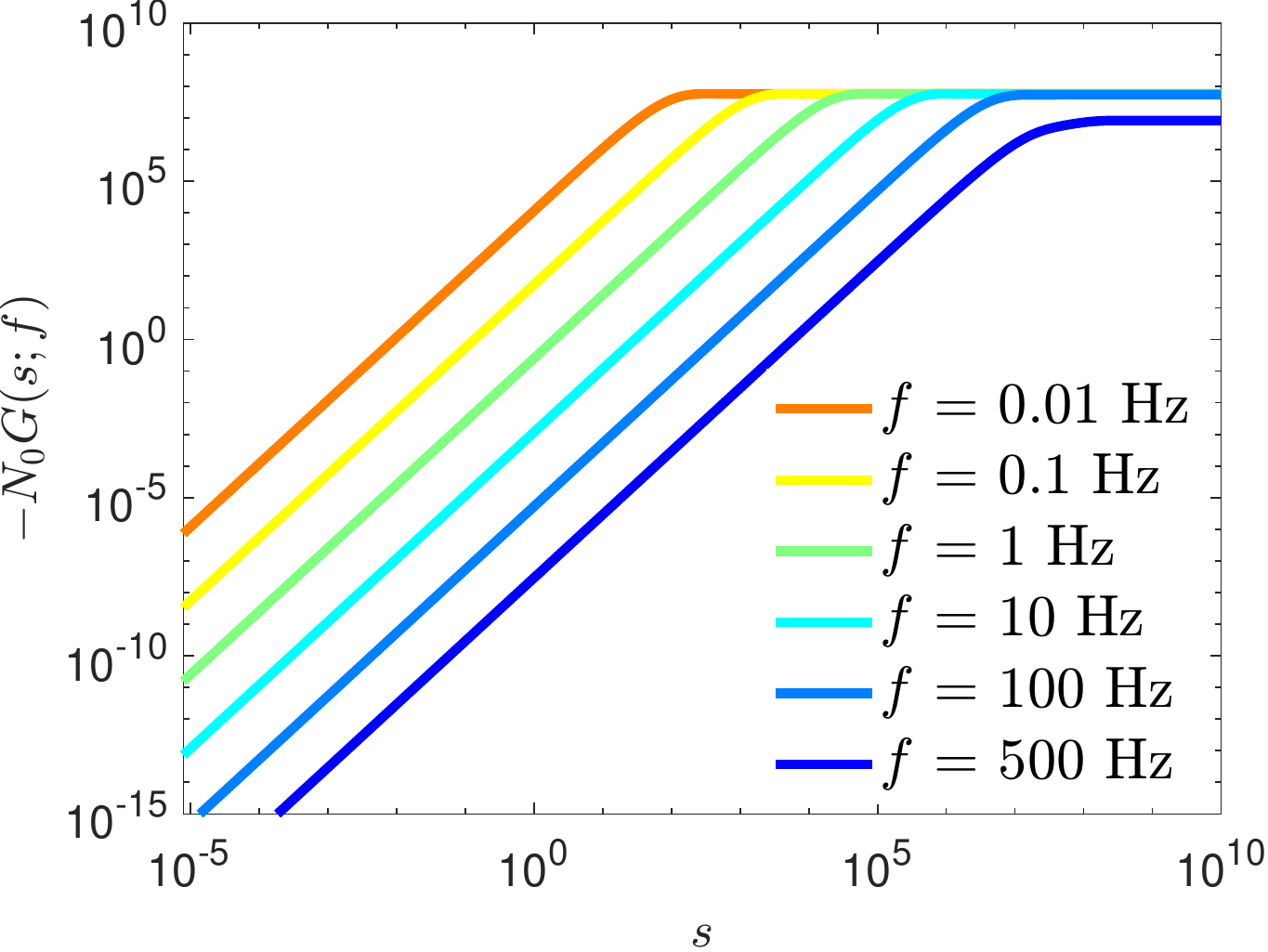}
    \caption{The function $-N_0G(s;f) > 0$, from equation \eqref{eq:ToyGsf1}, is shown for a range of observed frequencies $f$ as indicated in the legend. It scales as $-N_0 a(f) s^2$ at small $s$ and asymptotes to a $T$-dependent constant at large $s$ (see text for details). An observation time $T=1\yr$ is assumed for illustration.}
    \label{fig:Gsf}
\end{figure}

Similarly, the number density $\dd N/\dd f$ of overlapping sources is given by
\begin{align}
\label{eqn:dNdf}
    \frac{\dd N}{\dd f} = &\;N_0 \int_0^{r_0}\!\dd r\,r^2\int\!\dd^2\nvh\int\!\dd\cos\imath\,\int\!\dd M_c\int\!\dd t_* \nonumber \\
    &\times \frac{3}{8\pi r_0^3}\,\phi(t_*)\,\phi(M_c)\, \delta^D\big(f-f_o(\eta_0,r,M_c,t_*)\big) \nonumber \\
    =&\; \frac{4\pi N_0}{V_0} \int_0^{r_0}\!\dd r\,r^2\int\!\dd M_c \nonumber \\
    &\; \times\bigg\lvert\frac{\dd t_*}{\dd f}\bigg\lvert \,\phi\big(t_*(f,\eta_0,r,M_c)\big)\,\phi(M_c)\;,
\end{align}
where the formation time $t_*(f,\eta_0,r,M_c)$ solves the implicit equation $f=f(\eta_0,r,M_c,t_*)$. Hence,
\be
    \bigg\lvert\frac{\dd t_*}{\dd f}\bigg\lvert = \frac{8}{3f}\,t_*(f,\eta_0,r,M_c)\;,
\ee
as long as the observed frequency is $f<f_\text{cut}$ (and zero otherwise). This shows that the power-law behavior $\dd N/\dd f \propto f^{-11/3}$ is encoded in the Jacobian $|\dd t_*/\dd f|$.

We plot $\frac{\mathrm{d}N}{\mathrm{d}f}$ in figure \ref{fig:dndf}.
Observe that the pronounced frequency dependence of the source number density $\frac{dN}{df}(f)\propto f^{-11/3}$ reflects the time dependence of the rate of change $\dot\omega_r$ of the orbital frequency of a single binary. This frequency scaling would be somewhat different, had one relaxed the assumption of an adiabatic sequence of quasi-circular orbits.

\begin{figure*}
    \centering
    \includegraphics[width = .32\textwidth]{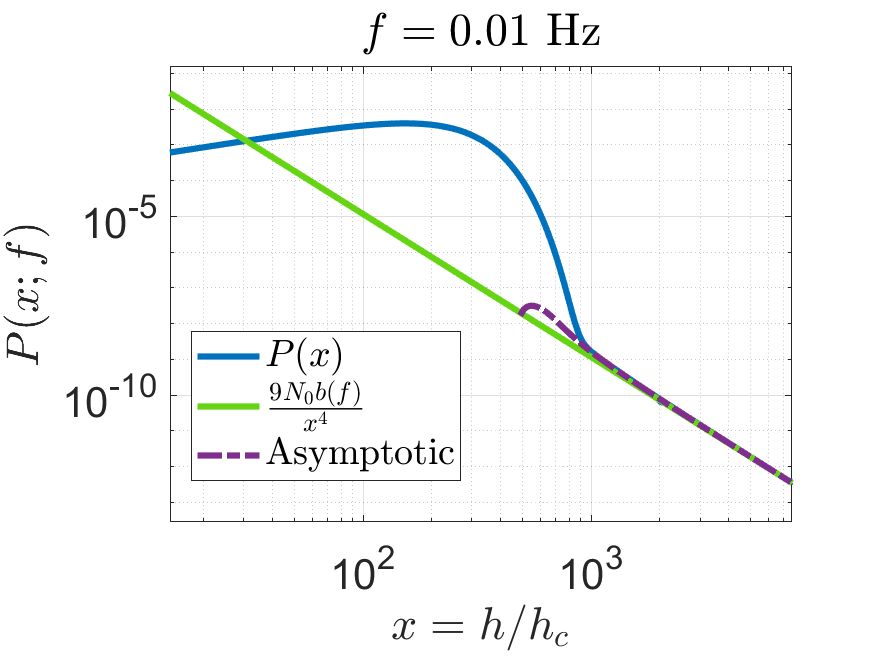}
    \includegraphics[width = .32\textwidth]{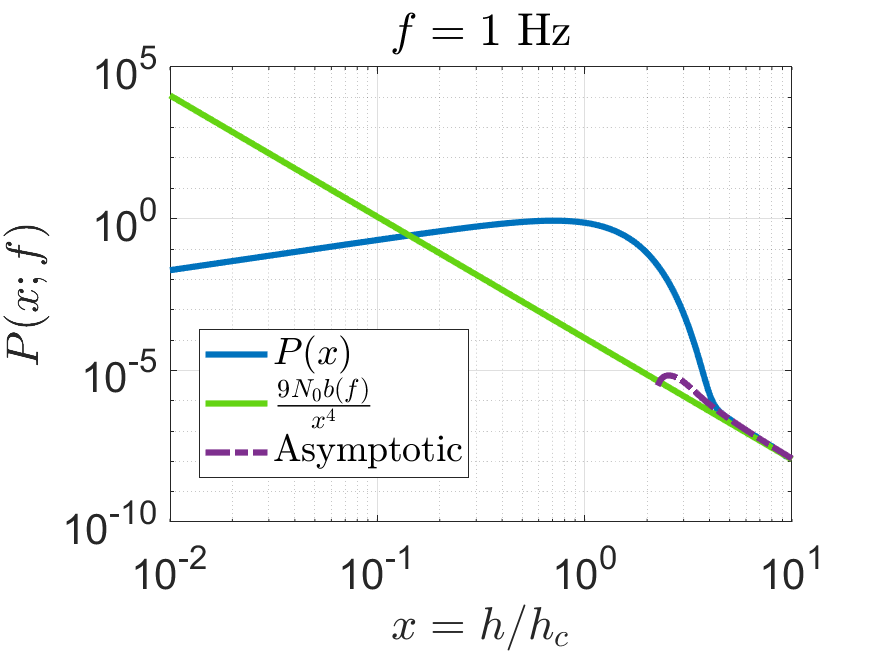}
    \includegraphics[width = .32\textwidth]{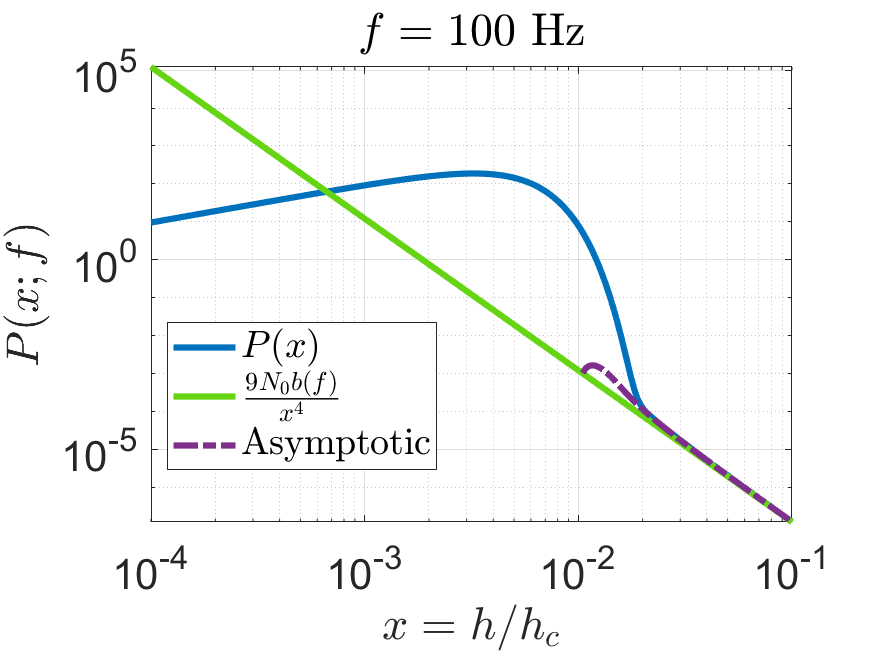}
    \caption{The probability density $P(x;f)$ as a function of the dimensionless ratio $x=h/h_c$ ($h=\tilde h_f$) for a few representative frequencies $f$. The blue curve is the exact prediction (obtained as the Hankel transform of $G(s;f)$), the violet dashed curve represents the asymptotic expansion \eqref{eqn: P_f asymptotic with Bessels}, and the green line indicates the power-law approximation Eq.~(\ref{eq:highstrainlaw}) valid at high strain. The amplitude $h$ of the observed DFT is normalized to $h_c=3.69\times 10^{-23}\Hz^{-1/2}$ in all panels.}
    \label{fig:Pxf simplified model}
\end{figure*}

\subsubsection{Generating function}

The generating function $G(s;f)$ reads
\begin{align}
    \label{eq:ToyGsf}
    G(s;f) &= \frac{3}{r_0^3}\int_0^{r_0}\!\dd r \,r^2\int\!\!\dd M_c\int\!\!\dd t_*\,\phi(t_*)\phi(M_c) \\
    & \qquad \times \frac{1}{8\pi}\int\!\dd^2\nvh\int\!\dd\cos\imath\,\bigg(J_0\Big(s\frac{|\tilde h_{f,\nvh}|}{h_c}\Big)-1\bigg). \nonumber
\end{align}
Extracting the factor of $\Pi_T(f)$ from the argument of the Bessel function, we can rewrite $G(s;f)$ as
\begin{align}
    \label{eq:ToyGsf1}
    G(s;f) &= \frac{3}{r_0^3}\int_0^{r_0}\!\!\dd r \,r^2\int\!\!\dd M_c\int\!\!\dd t_*\,\Pi_T(f)\,\phi(t_*)\,\phi(M_c) \\
    & \qquad \times \frac{1}{8\pi}\int\!\dd^2\nvh\int\!\dd\cos\imath\,\bigg(J_0\Big(s\frac{|\tilde h(f,\nvh)|}{h_c\sqrt{T}}\Big)-1\bigg)\;, \nonumber
\end{align}
where our approximation to the window function of the experiment is
\begin{equation}
\begin{split}
    \Pi_T(f) = \; & \Theta\big(f-f_o(\eta_0,r,M_c,t_*)\big) \\ &\times \Theta\big(f_o(\eta_0+T,r,M_c,t_*)-f\big) \;.
\end{split}
\end{equation}
In practice, $\Pi_T(f)$ implies a constraint on $t_*$ which we take advantage of to restrict the domain of the $t_*$-integration. Eq.~(\ref{eq:ToyGsf1}) implies that the $k$-th moment of the distribution, which is proportional to the $s^k$ term in the series expansion of $J_0$, decays as $T^{1-k/2}$ relative to the variance.

Having calculated $G$, one may now insert it into equation \eqref{eqn:P(|h|) bessel} and calculate $P(h;f)$. We do so numerically in figure \ref{fig:Pxf simplified model}, and also plot the analytic asymptotic expansion \eqref{eqn: P_f asymptotic with Bessels} and the power-law tail \eqref{eq:highstrainlaw}.

The characteristic function $G(s;f)$ and the 1-point distribution $P(x;f)$ displayed in figures \ref{fig:Gsf} and \ref{fig:Pxf simplified model} are computed for an observation time $T=1\yr$. They are shown for different frequencies as indicated in the panels. A unique characteristic strain $h_c$ is used to define $x=h/h_c$ throughout the panels; we chose $h_c=3.69\cdot 10^{-23} \Hz^{-1/2}$ in this sub-section to match the rms variance of the Fourier mode amplitude at $f=1\Hz$, so that $P(x;f)$ peaks around $x=1$ for $f=1\Hz$.

For $s\gg 1$, $-G(s;f)$ converges to (minus) the same effective volume of the parameter space (which is proportional to $T$) for all frequencies except $f=500\Hz$, where $-G(s;f)$ is lower due to the rapid decline in the number of contributing sources. For $s\ll 1$, $G(s;f)$ admits the series expansion $-a(f) s^2 + b(f) |s|^3$. The frequency-dependent coefficients scale like $a(f)\propto f^{-7/3}$ and $b(f)\propto f^{-7/2}$ and can be accurately determined as explained in Appendix \S\ref{app:G approximation}. We exploit this to mitigate numerical noise when $s\ll 1$ and improve the computation of $P(x;f)$, substituting $-as^2+b\abs{s}^3$ for $G$ at $s\ll 1$, when evaluating the Hankel transform \eqref{eqn:P(|h|) bessel}, both here an in \S\ref{sec:realisticSGWB}.

Fig.~\ref{fig:Pxf simplified model} shows the distribution $P(x;f)$ computed as the Hankel transform of $G(s;f)$ for three different frequencies as indicated in the panels. Due to the very large $N_0\gg 1$, a (Gaussian) Rayleigh distribution plus the power-law tail approximation Eq.~(\ref{eq:highstrainlaw}) is an excellent approximation for the observed frequencies shown here. Before proceeding to generalize this to a realistic model, let us comment on how to derive $\Omega_{\rm gw}$.

\subsubsection{GW energy spectrum}

Using equations (\ref{eq:singlesourceDFT1}) and (\ref{eq:singlesourceDFT}), the second moment $\langle|\tilde h_f|^2\rangle$ of the Fourier amplitudes can be analogously expressed as
\begin{align}
    \langle |\tilde h_f|^2 \rangle &= \frac{3N_0}{r_0^3}
    \int_0^{r_0}\!\dd r \,r^2\int\!\dd M_c\int\!\dd t_*\,\frac{\Pi_T(f)}{T}\,\phi(t_*)\phi(M_c)\nonumber \\
    & \qquad \times \frac{1}{8\pi}\int\!\dd^2\nvh\int\!\dd\cos\imath\,|\tilde h(f,\nvh)|^2 \label{eq:averagehf2} \;.
\end{align}
The angular average of the single-source amplitude squared $|\tilde h(f,\nvh)|^2$ returns \citep[see][]{Maggiore}
\begin{align}
    \frac{1}{8\pi}\int\!\dd^2\nvh\int\!\dd\cos\imath\,|\tilde h(f,\nvh)|^2 &= h_0^2(f)\, \big\langle \big\lvert Q(\vartheta,\varphi,i)\big\lvert^2\big\rangle_{\nvh,\imath}\nonumber \\ &= \frac{2}{5}\, F \, h_0^2(f) \;, \label{eq:averageQ}
\end{align}
so that the second moment becomes
\begin{align}
    \langle |\tilde h_f|^2 \rangle &= \frac{6 F N_0}{5r_0^3}
    \int_0^{r_0}\!\dd r \,r^2\int\!\dd M_c\int\!\dd t_* \nonumber \\
    &\qquad \times \frac{\Pi_T(f)}{T}\,\phi(t_*)\,\phi(M_c)\, h_0^2(f) \;.
\end{align}
Substituting this result into Eq.~(\ref{eq:FSh}), we can extract an expression for $S_h(f)$ and recast the GW energy spectrum $\ogw(f)$ into the form (for $\iota = 0$)
\begin{align}
    \label{eq:ToyOmegagw}
    \ogw(f) &= \frac{16\pi^2}{5r_0^3 H_0^2}\, N_0\, f^3\int_0^{r_0}\!\dd r\, r^2 \int\!\dd M_c\int\!\dd t_* \\
    &\qquad \times \frac{\Pi_T(f)}{T} \,\phi(t_*)\, \phi(M_c) \, h_0^2(f)
    \nonumber \;.
\end{align}
The rectangular window $\Pi_T(f)$ depends on the model parameters and, therefore, cannot be taken out of the integral. However, since $\Pi_T\propto T$, both $S_h(f)$ and $\ogw(f)$ are independent of $T$, in the limit $T\to 0$. For an observation time $T\ll \eta_0$, we can write
\be
    f_o(\eta_0+T,r,M_c,t_*)\approx
    f_o(\eta_0,r,M_c,t_*) + T\frac{\dd f_o}{\dd\eta}\bigg\lvert_{\eta=\eta_0}\;,
\ee
which shows that
\begin{align}
    \lim_{T\to 0}\frac{\Pi_T(f)}{T} &= \bigg\lvert\frac{\dd f_o}{\dd\eta}\bigg\lvert_{\eta=\eta_0}
    \delta^\text{D}\!\big(f-f_o(\eta_0,r,M_c,t_*)\big) \\
    &= \bigg\lvert\frac{\dd t_*}{\dd\eta}\bigg\lvert_{\eta=\eta_0}
    \delta^\text{D}\!\big(t_*-t_*(f,\eta_0,r,M_c)\big) \nonumber \;.
\end{align}
Substituting this relation into Eq.~(\ref{eq:ToyOmegagw}) and taking advantage of $\dd t_*/\dd\eta=(1+z)^{-1}$, we arrive at
\begin{align}
    \ogw(f) &= \frac{16\pi^2}{5r_0^3 H_0^2}\, N_0\, f^3\int_0^{r_0}\!\dd r\, r^2 \int\!\dd M_c\, (1+z)^{-1} \nonumber \\
    &\qquad \times \phi\big(t_*(f,\eta_0,r,M_c)\big)\, \phi(M_c) \, h_0^2(f)
    \;. \label{eq:ToyOmegagw1}
\end{align}
The shape of $\ogw(f)$ reflects the dependence of the Fourier amplitudes, Eq.~\ref{eq:FTamplitudes}, on frequency. For the single population model considered here, the power-law behavior $\ogw(f)\propto f^{2/3}$ at low $f$ is followed by a mild rise and a sharp suppression at high frequencies. We plot $\Omega_{\rm gw}(f)$ in figure \ref{fig:Omegagw} for the more realistic model we now consider. We refer the readers to Appendix \ref{appendix:comparison with Phinney} for a comparison of equation \eqref{eq:ToyOmegagw1} with other expressions in the literature.

\subsection{Un-subtracted GW signal of compact stellar remnants}
\label{sec:realisticSGWB}

We now turn to the unsubtracted GW signal arising from all the mergers of neutron stars and stellar-mass black holes produced by the core collapse of massive stars. We do not distinguish between the different types of compact binaries because unresolved signals eventually comprise the SGWB, and as such, it is impossible to determine which types of binary the SGWB comes from.
This assumption only changes the overall amplitude of the signal, which is not the focus here, but the dependence of $P(h)$ on $h$ will not change.

The joint distribution function $\phi(\vr,\vxi)$ of the source co-moving position $\vr=(r,\vartheta,\varphi)$ and intrinsic properties $\vxi=(t_*,T_*,m_1,m_2,\imath,\dots,\mu)$ follows the `reference model' of \cite{Cusinetal2019}. Details can be found in Appendix \ref{app:galaxymodel}. 
In particular, the initial period $T_*$ follows \"{O}pik's law, i.e. it is uniform in $\ln T_*$. As this is proportional to $\ln \tau_0$, one can instead switch from $T_*$ to $\tau_0$ as a model parameter, with the measure uniform in $\ln \tau_0$. Then, inequality \eqref{ineq:heaviside condition} may be analytically integrated, as follows: upon changing from conformal to cosmic time, and by requiring that $T \ll t_0$, we find
\be
    t_{0,\textrm{ret}}(r) \leq \tau_0 - \mathfrak{t}_s + t_* \leq t_{0,\textrm{ret}}(r) + T \;.
\ee
Consequently, the integral over $\tau_0$ is just
\be\label{eqn: pi T analyic integration}
    \ln \left(\frac{\tau_{\max}}{\tau_{\min}}\right),
\ee
where
\begin{align}
    \tau_{\min} & \equiv \max\set{\tau_0(M_c,T_{\min}),t_{0,\textrm{ret}}(r) + \mathfrak{t}_s(f,r,\boldsymbol{\xi})\! - \!t_*} \\
    \tau_{\max} & \equiv \min\set{\tau_0(M_c,T_{\max}),t_{0,\textrm{ret}}(r) + \mathfrak{t}_s(f,r,\boldsymbol{\xi}) \!- \!t_* \!+\! T} \nonumber \;,
\end{align}
where $T_{\max}$ and $T_{\min}$ are the maximum and minimum allowed initial periods in the model respectively.
In the (realistic) regime of small observation times $T\ll t_0$, the double-Heaviside condition may be converted into
\begin{equation}
    \Pi_T \approx T \delta^\text{D}\!\left(\tau_0 - \mathfrak{t}_s(f,r,\boldsymbol{\xi}) + t_* -t_{0,\textrm{ret}}\right) \;.
\end{equation}

We have also added gravitational lensing by intervening matter. Gravitational lensing alters the amplitude of the detected GW strain by a factor of $\sqrt{\mu}$ where $\mu$ is the magnification. Under the assumption of Poisson distributed sources, the lens and the source are uncorrelated. Therefore, for a given $\mu$, we only have to re-scale the Fourier amplitudes according to $|\tilde h(f)| \mapsto \sqrt{\mu}|\tilde h(f)|$. The distribution $\phi(\mu|z(r))$ of lensing magnification given in Appendix \ref{app:lensing} is a function of the source redshift.
Summarizing, $G(s;f)$ for the detailed model is given by
\begin{widetext}
\begin{equation}\label{eqn:G}
\begin{aligned}
  G(s;f) & = 3\int_{r_{\min}/r_0}^{R/r_0}\! \mathrm{d}\tau\, \tau^2\int_{\mathcal{S}^2}\! \mathrm{d}\Omega \int_{0}^{\pi}\!\mathrm{d}\iota \sin \iota \iint\! \mathrm{d}m_1\mathrm{d}m_2 \int\! \mathrm{d}T_* \int\! \mathrm{d}t_* \int\! \mathrm{d}M_\textrm{gal}\int_0^{\infty}\!\mathrm{d}\mu\, \phi(t_*,m_1,m_2,T_*,M_\textrm{gal})\phi(\mu|z(r)) \\ &
  \qquad\times \left[J_0\left(s\frac{h_{\rm eff}(f,r,\boldsymbol{\xi})}{h_c}\sqrt{\frac{\mu}{T}}\right) - 1\right]\,\Theta\big(f-f_o(\eta_0,r,\vxi)\big)\,\Theta\big(f_o(\eta_0+T,r,\vxi)-f\big).
\end{aligned}
\end{equation}
Upon simplification with equation \eqref{eqn: pi T analyic integration} and \"{O}pik's law for $T_*$, it becomes
\begin{equation}
\begin{aligned}
  G(s;f) & = \frac{3}{4}\int_{r_{\min}/r_0}^{R/r_0}\! \mathrm{d}\tau\, \tau^2\int_{\mathcal{S}^2}\! \mathrm{d}\Omega \int_{0}^{\pi}\!\mathrm{d}\iota \sin \iota \iint\! \mathrm{d}m_1\mathrm{d}m_2 \int\! \mathrm{d}t_* \int\! \mathrm{d}M_\textrm{gal}\int_0^{\infty}\!\mathrm{d}\mu\, \phi(t_*,m_1,m_2,M_\textrm{gal})\phi(\mu|z(r)) \\ &
  \qquad\times \left[J_0\left(s\frac{h_{\rm eff}(f,r,\boldsymbol{\xi})}{h_c}\sqrt{\frac{\mu}{T}}\right) - 1\right]\ln \left(\frac{\tau_{\max}}{\tau_{\min}}\right)\,\left[\ln\left(\frac{a_{\max}}{a_{\min}}\right)\right]^{-1}.
\end{aligned}
\end{equation}
Finally, the requirement $R_\text{merger}=1000\gpc^{-3}\yr^{-1}$ consistent with the Advanced {\small LIGO} and {\small VIRGO} O3 data \cite{LIGOvirgo2021PRX} yields $N_0 = 7.33 \times 10^{17}$, via equation \eqref{eq:mergerrate}, and we take $h_c = \mathbb{E}\left[|\tilde{h}(f=1\textrm{ Hz})|\right] = 5.274\times 10^{-31}~\textrm{yr}^{1/2}$.
\end{widetext}

\begin{figure}
    \centering
    \includegraphics[width = .45\textwidth]{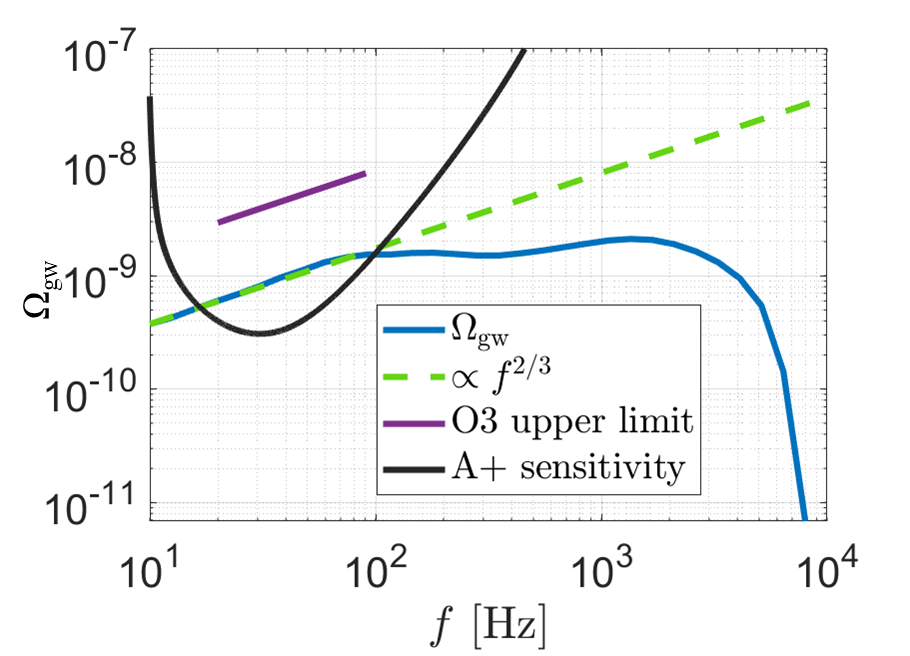}
    \caption{The GW energy density as a function of frequency for the unsubtracted signal produced by all stellar compact remnants in the Universe. The low- and high-frequency bumps correspond to binary BH and NS mergers, respectively. The upper limit inferred from the Advanced {\small LIGO}-Virgo-{\small KAGRA} O3 data is indicated on the figure. We also plot the $A+$ sensitivity curve for an observation time $T=1\yr$ and a frequency resolution $\Delta f=0.25\Hz$.}
    \label{fig:Omegagw}
\end{figure}

Fig.~\ref{fig:Omegagw} shows the GW energy density of the unsubtracted GW strain as a function of the measured frequency. The two local maxima at $f\sim 100$ and $\sim 10^3\Hz$ correspond to binary BH and NS mergers, respectively. Note that the energy spectrum significantly deviates from the $f^{2/3}$ scaling at frequencies $f\gtrsim 100\Hz$. The current upper limit on the energy density of this background inferred from the O3 run, $\ogw\leq 3.4\times 10^{-9}$ at $f=25\Hz$ (for a $f^{2/3}$ spectrum in the range $20-90\Hz$) \cite{LIGOVIRGOSGWB}, is indicated in the figure along with the sensitivity of a single A+ detector with observation time $T=1\yr$ and frequency resolution $\Delta f=25\Hz$ \cite{ThraneRomano2013,Barsottietal2018a,Barsottietal2018b}~\footnote{The noise PSD is available at \url{https://dcc.ligo.org/LIGO-T1800044/public}.}

In figure \ref{fig:Pxf}, we show the corresponding $P(x;f)$ at observed frequency $f=5, 50$, $500$ and $2000\Hz$, with and without the effect of lensing. As before, it asymptotes to a power-law tail $9N_0b(f)x^{-4}$. For values of $h$ less than the threshold above which the power-law dominates, the distribution is very close to Gaussian with sub-percent deviations from a Rayleigh distribution (we found that deviations larger than a percent are obtained for $N_0\lesssim 10^{15}$). Gravitational lensing induces a percent level shift of the distributions to larger strains. Although it can dramatically enhance the source brightness on rare occasions \citep[see the discussion in][]{dai/venumadhav/sigurdson:2017}, it does not affect the $h^{-4}$ slope of the power-law tail, which reflects the $1/r$ dependence of the signal.

\subsection{SGWB}
\label{sec:confusionnoise}

We have thus far computed the distribution $P(|\tilde h_f|)$ of Fourier mode amplitudes produced by all the sources giving rise to the astrophysical GW signal of compact binary mergers. In practice, bright mergers will be identified and removed from the raw GW strain \citep{cutler/harms:2006,Timpanoetal2006,harms/etal:2008,Regimbau:2009rk,cornish/robson:2017,Pieroni:2020rob}. The remaining, unresolved binaries act as an effective noise source which diminishes as data is acquired and more bright sources are removed. The resulting distribution, which we denote $P_r(h;f)$, will characterize the so-called SGWB of unresolved sources. Since the identification, modelling and subsequent removal of bright sources are detector-dependent, we shall consider here the following simplified implementation, similar to that of \cite{Ginatetal2020}.

\begin{figure*}
    \centering
    \includegraphics[width = .46\textwidth]{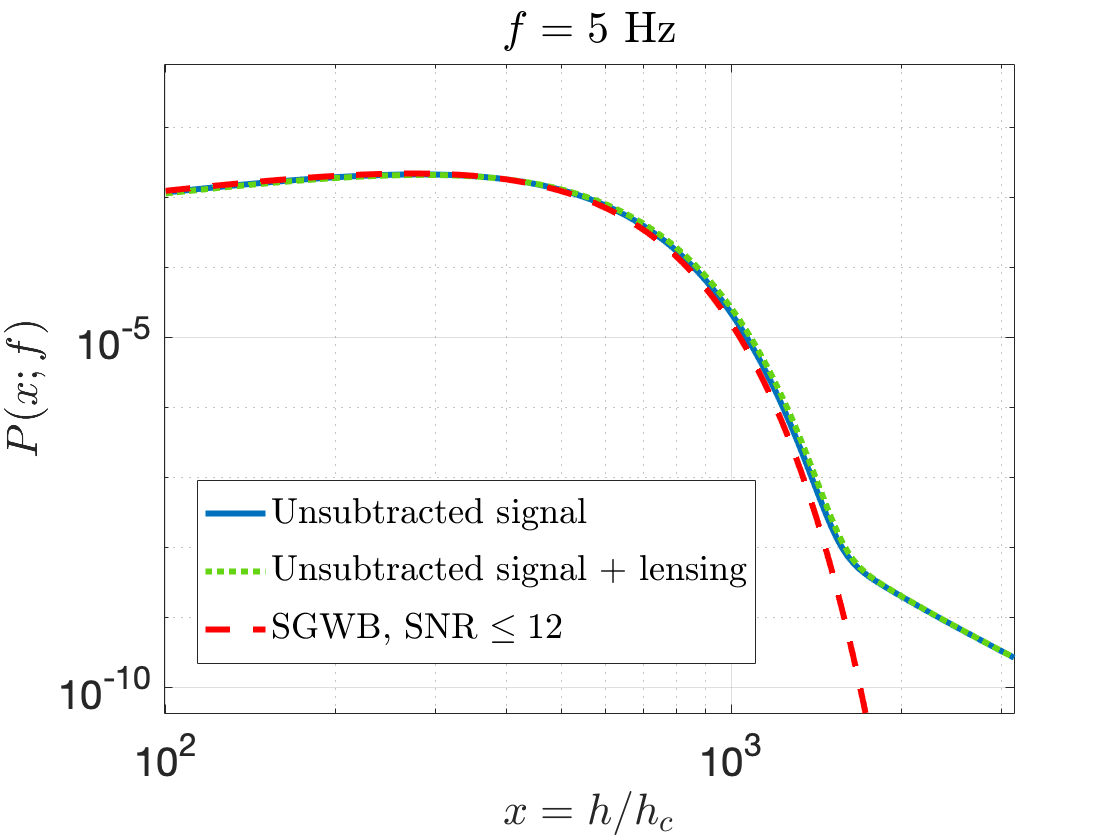}
    \includegraphics[width = .46\textwidth]{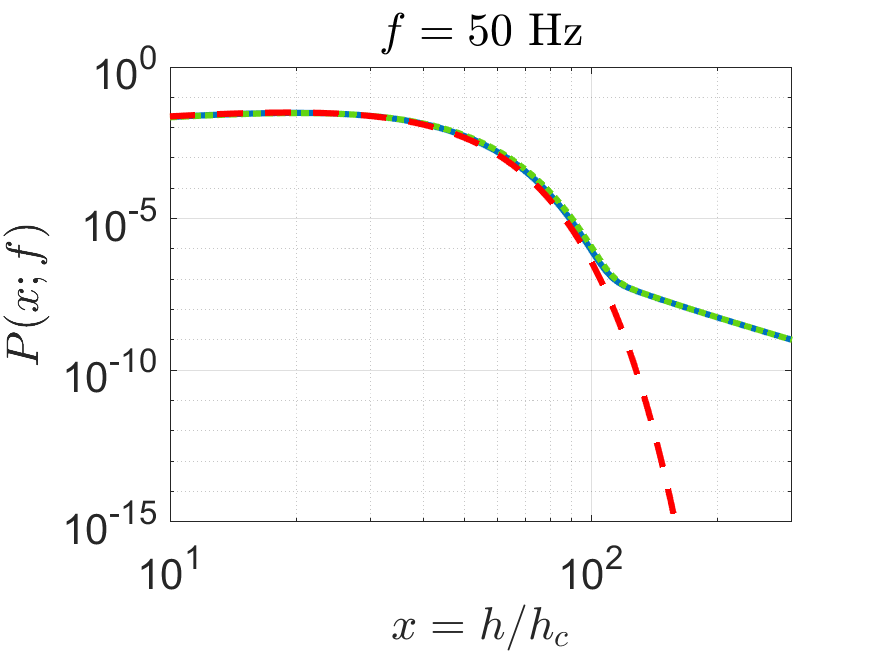}
    \includegraphics[width = .46\textwidth]{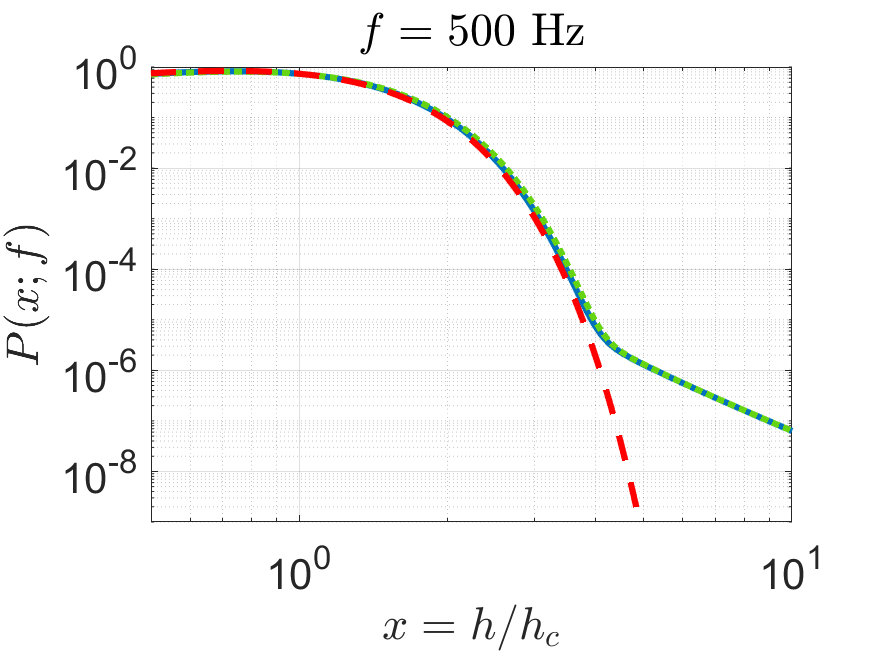}
    \includegraphics[width = .46\textwidth]{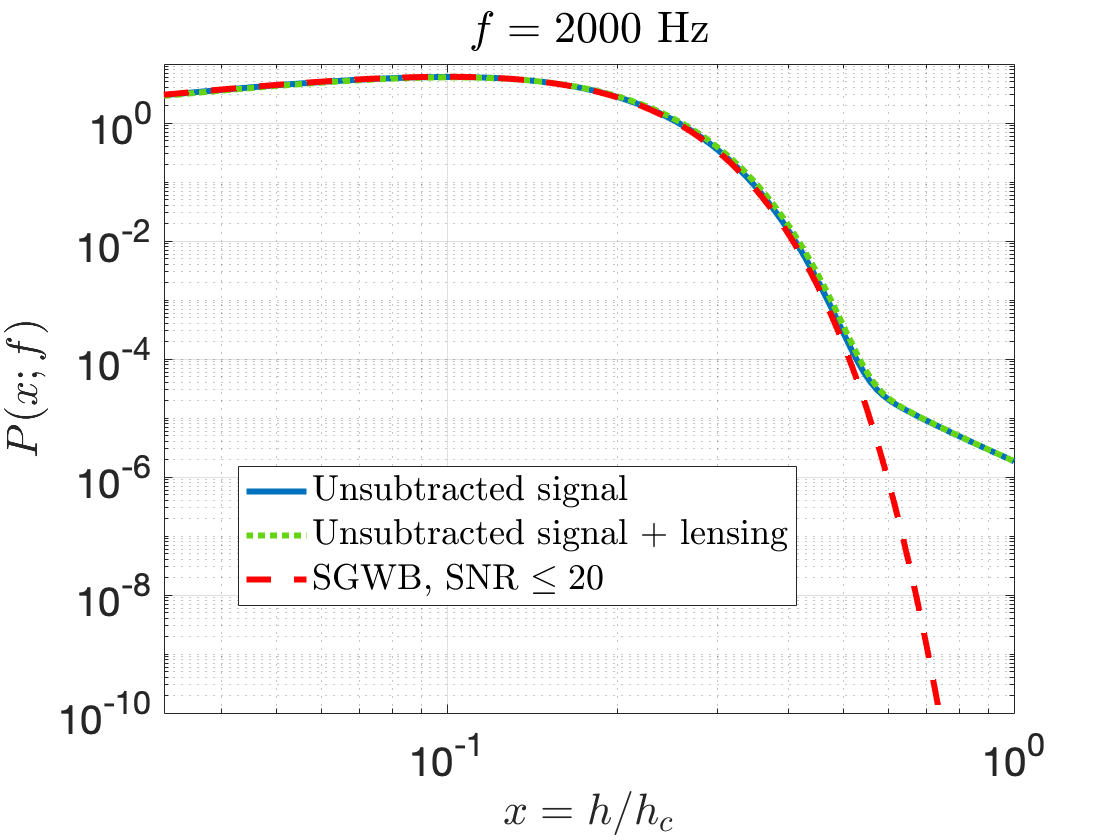}
    \caption{Distribution of discrete Fourier amplitudes $h=|\tilde h_f|$ as a function of the dimensionless ratio $x=h/h_c$. The strain normalization is $h_c=5.274\times10^{-31}~\textrm{yr}^{1/2}$ and the observation time set to $T=1\yr$. The solid (blue) and dotted (green) curves show the distribution $P(x;f)$ of the unsubtracted signal produced by all the cosmological compact binaries without and with lensing. The dashed (red) curve is the distribution $P_r(x;f)$ of the SGWB originating from unresolved sources with SNR$\leq 12$ assuming a single detector of A+ sensitivity and an observation time $T=1\yr$ (see text for details), except for $2000$, which has a threshold of $20$, just to show how varying the threshold affects $P_r$.}
    \label{fig:Pxf}
\end{figure*}

The signal-to-noise ratio (SNR) of the detector for a single binary event is
\begin{align}
    \text{SNR}^2 &= 4 \sum_n \frac{|\tilde h_{f_n}|^2}{S_n(f)} \\
    &= 4 T \int\!\dd f\, \frac{|\tilde h(f)|^2}{S_n(f)}\,\frac{\Pi_T(f)}{T} \nonumber \\
    &= 4 \int\!\dd f\, \frac{|\tilde h(f)|^2}{S_n(f)}\,\Pi_T(f) \nonumber
\end{align}
where $S_n(f)$ is the detector noise PSD. The factor of $\Pi_T(f)$ takes into account that a compact stellar binary emits GWs with frequency rising in time, until its components coalescence.

To avoid dealing with the factor of $\Pi_T(f)$ on a binary-by-binary case, we assume that the detector records all the merger events while the frequency $f^\textrm{obs}_\textrm{gw}$ of the gravitational waves lies between the detector's $f_{\min}$ and $f_{\max}$, independently of the value of $T$
~\footnote{This is a good approximation so long as the detector records the coalescing binary when it enters its frequency band. It fails when e.g. the binary already is in the detector band when the latter is turned on.}. Therefore, we approximate the SNR by
\be
    \text{SNR}^2 \simeq 4 \int_{\fmin}^{\fmax}\!\dd f\, \frac{|\tilde h(f)|^2}{S_n(f)} \;.
\ee
Furthermore, on using the Newtonian wave-form for the SNR computation (adopting the template \ref{eq:FTamplitudes} does not make a significant difference), the SNR depends on a common factor of
\be
    I = \int_{\fmin}^{1700\Hz}\!\dd f\,\frac{f^{-1/3}}{f^2S_n(f)}
\ee
Owing to the $f^{-7/3}$ power in the integrand, changing the upper limit of $I$ from the maximum possible value of $\fmax\sim 2f_\text{ISCO}$ in the model used here, i.e. $\sim 1700\Hz$ (corresponding to the merger of a NS binary), to its minimum, has a small effect on the value of $I$. Thus, requiring that the source SNR exceed a given threshold $\rho$ amounts to the condition
\begin{equation}
\label{eqn: bright condition}
\begin{split}
    \frac{(M_c (z+1))^{5/3}}{d_L^2} \geq &\;\frac{3\pi^{4/3}c^3 \rho^2}{G^{5/3}I}\\ \approx &\; 3\rho^2\times 10^{-7} M_\odot^{5/3} ~\textrm{Mpc}^{-2}\;.
\end{split}
\end{equation}
Hereafter, we adopt the conservative detection threshold of $\rho=12$ \cite{abbott_prospects_2018}. If this inequality is satisfied, the source is deemed \emph{bright}, and its signal is removed from the data provided that its time to coalescence $\mathfrak{t}$ is smaller than $T$ (so that it merges during a $T=1\yr$ observational run). In plain words, we remove the entire source's contribution to $G(s;f)$.
In practice, the bound on the SNR may be formally expressed as a bound on a function of the source parameters.
The latter is then inserted as a Heaviside function into the integrand defining $G(s;f)$, thereby ensuring that the condition \eqref{eqn: bright condition} is not satisfied by the sources making up the confusion background.
In Fig.~\ref{fig:Pxf}, the distribution of the resulting SGWB is shown as the dashed (red) curve for the observed frequencies quoted in the figure.

It is clear that $P_r(h;f)$ differs significantly from $P(h;f)$ at high strain when bright sources are removed: beyond a certain threshold $h_{\rm cut}$, the only possibility to have a residual confusion noise $h \gtrsim h_{\rm cut}$ is if $h$ is composed of constructively-interfering weak signals from many sources, each of which is too weak to be individually resolved. Clearly, the probability for this event is exponentially small in $h$.

As shown in \cite{Ginatetal2020}, the removal of bright sources regularizes $G(s;f)$ in the time domain, and it does so here, too. The shape of the exponential decline turns out to be mostly sensitive to the analytic continuation of $G(s;f)$ into the complex plane at large values of $\Im s \sim \ln h/h_c$ (see appendix \ref{app:bright sources}). This implies, \emph{inter alia}, that a direct computation of this decline with a direct, numerical computation of $G(s;f)$ is quite difficult because the Bessel function $J_0$ both oscillates and grows exponentially; equivalently, $G(s;f)$ on the real axis must be evaluated with extremely high accuracy, in order for the analytic continuation -- i.e. for its Hankel transform at large $h$ -- to be accurate.

For an observation time $T=1\yr$, we find that the `intermediate' $h$ expansion of appendix \ref{app:bright sources} applies, for all the frequencies we consider. Furthermore, when the bright sources resolved during this $1\yr$ observational run are removed from the entire data, the quadratic term $as^2$ in $G(s;f)$ dominates and, thereby, $P_r(h;f)$ is essentially a Rayleigh distribution (as is apparent from Fig.~\ref{fig:Pxf}). Technically, this originates from the fact that the critical point is $s = \mathrm{i}h_{\max}/(2N_0 ah_c)$ and satisfies $\abs{ds^2},\abs{bs} \ll a$ ($b$ and $d$ are the 3rd and 4th moments) for all the frequencies shown in Fig.~\ref{fig:Pxf}

The expressions and techniques used here are general and apply to any integration time and observed frequency. Therefore, for the sake of completeness, let us briefly comment on what happens when the aforementioned (intermediate $s$) solution fails. In Appendix \S\ref{app:bright sources} we derive an asymptotic form for $P_r(h;f)$ at arbitrarily large values of $h$, which reads
\begin{widetext}
\begin{equation}\label{eqn:P_r asymptotic}
    P_r(h;f) \sim \sqrt{\frac{h}{h_{\max}^3}} \exp\left(-\frac{h}{h_{\max}} -\frac{3h}{2h_{\max}}\abs{W_{-1}\left[-\frac{2h_{\max}}{3}\left(\frac{N_0h_{\max}C}{h}\right)^{2/3}\right]}\right)\left(\frac{\pi^2}{4} + \ln^2\frac{h}{h_{\max}}\right)^{1/2},
\end{equation}
where $W_{-1}$ denotes Lambert's $W$-function, whereas $h_{\max}$ and $C$ are coefficients that can be calculated directly given a threshold SNR $\rho$. Their explicit expressions can be found in \S\ref{app:bright sources}. Concretely, $h_{\max}$ is the threshold strain amplitude $h_0(f)/\sqrt{T}$ at a given observed frequency $f$ above which the SNR condition (\ref{eqn: bright condition} is satisfied.
\end{widetext}

\begin{figure}
    \centering
    \includegraphics[width=0.45\textwidth]{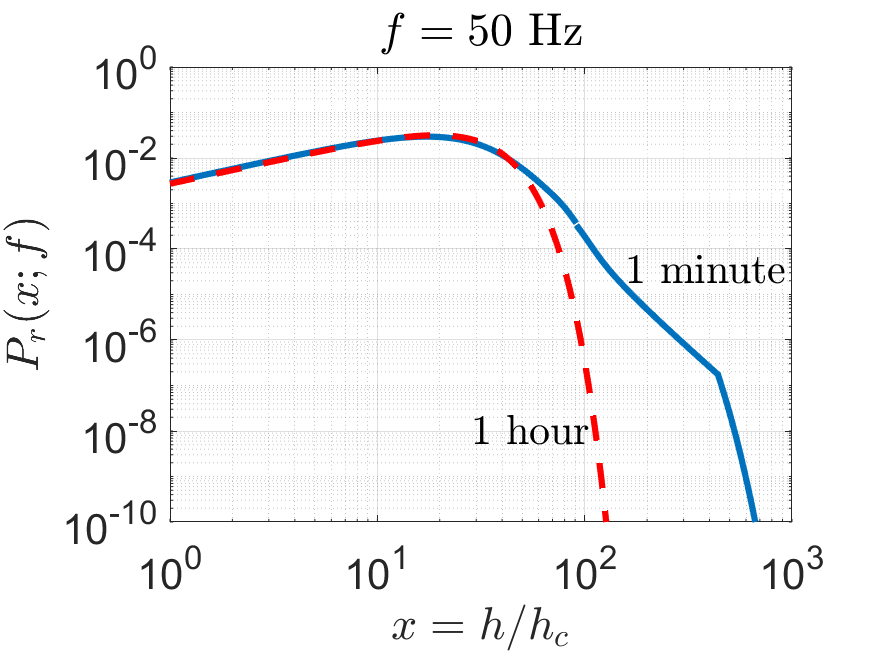}
    \caption{The SGWB at $50$ Hertz when bright sources are subtracted for an observation time of 1 minute or 1 hour.}
    \label{fig:1 minutes observation time}
\end{figure}

For short observation times $T\ll\yr$ (or low detector sensitivity), not all of the power-law tail is resolved, and the effective cut-off strain becomes larger. Then, $P_r(h;f)$ follows $P(h;f)$ up to a cut-off $h_{\max}$, when it assumes the form \eqref{eqn:P_r asymptotic}. This is illustrated in Fig.~\ref{fig:1 minutes observation time}. To compute $P_r(h;f)$ in this figure, we evaluated the Hankel transform up to the point where the asymptotic \eqref{eqn: P_f asymptotic with Bessels} became accurate, and then plotted it until it became larger than \eqref{eqn:P_r asymptotic}, which is where the latter becomes the accurate expression for $P_r$ (Evaluating the Hankel transform directly over the entire range of strains was numerically unstable). For the case of $T = 1$ minute, we calculated $h_{\max} = 244.3h_c$; one can see that the transition from the $h^{-4}$ occurs close to $h_{\max}$, strengthening the physical intuitive reasoning that the former power-law comes from the brightest, single, unresolved source within the observation run. For $T = 1$ hour, the situation already becomes similar to the longer observation times considered above.

\begin{figure}
    \centering
    \includegraphics[width=0.45\textwidth]{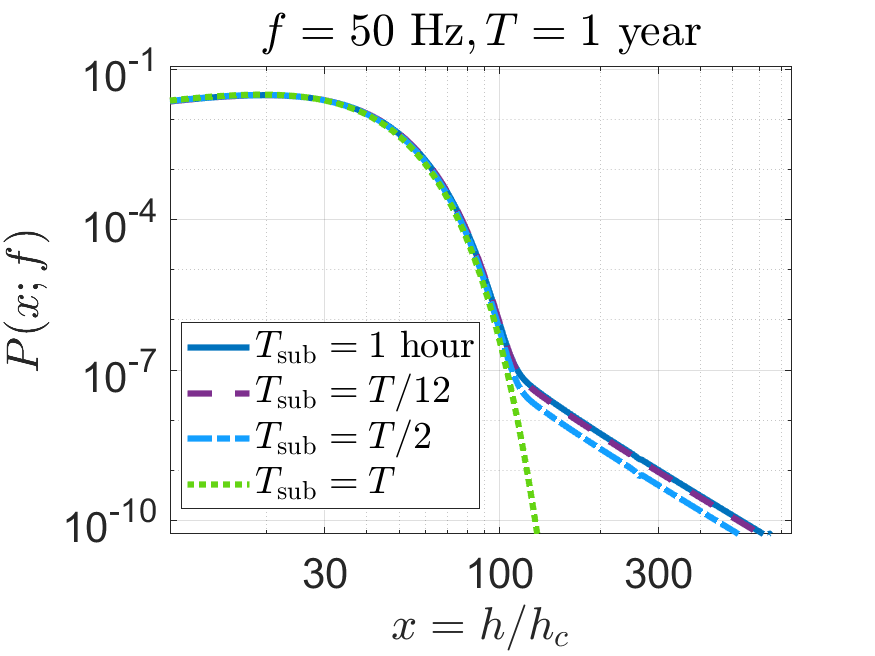}
    \includegraphics[width=0.45\textwidth]{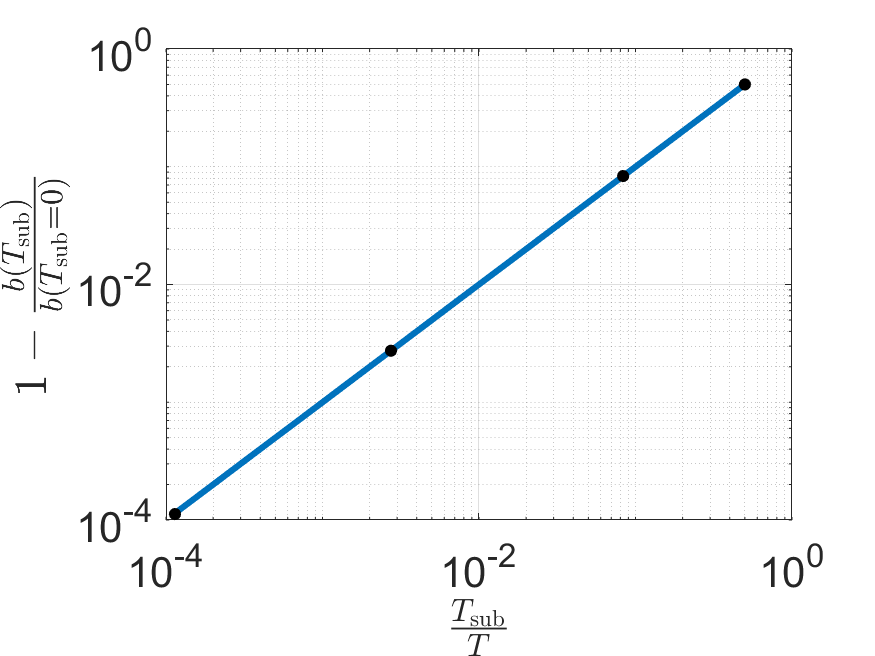}
    \caption{{\it Top panel}: Distribution of the SGWB for various choices of $T_{\rm sub}$ at fixed observation time $T = 1\yr$. Results are shown at a frequency of $50\Hz$ for illustration. The amplitude of the $h^{-4}$ tail decreases with increasing $T_{\rm sub}/T$, as in equation \eqref{eqn: tail amplitude as function of T sub}, and vanishes for $T_{\rm sub} = T$, leaving an almost-Rayleigh distribution. {\it Bottom panel}: a validation of equation \eqref{eqn: tail amplitude as function of T sub}, obtained by comparing it with the numerically evaluated values of $b$ (using equation \eqref{eq:bf}).}
    \label{fig:subtraction times}
\end{figure}

A different approach consists of fixing the observation time $T = 1$ year and defining another parameter, $T_{\rm sub} \leq T$, such that only mergers occurring during the final period of duration $T_{\rm sub}$ are removed, and the rest are kept as unresolved. Of course, the position of the $T_{\rm sub}$ interval within the observation period does not change anything. This is a different situation from above, because now bright sources are still allowed to exist, and therefore the $|s|^3$ singularity is still present for all $T_{\rm sub} < T$. In fact, by stationarity of the SGWB, its amplitude is given by
\begin{equation}\label{eqn: tail amplitude as function of T sub}
    1-\frac{b(T_{\rm sub})}{b(T_{\rm sub} = 0)} = \frac{T_{\rm sub}}{T},
\end{equation}
with 
\begin{equation}
    \lim_{T_{\rm sub} \nearrow T} b = 0,
\end{equation}
thereby removing the singularity continuously. On the other hand, $a$ goes to a finite value as $T_{\rm sub} \to T$, and is generally insensitive to it, as it is insensitive to $T$. $P_r$ is plotted in figure \ref{fig:subtraction times} for various choices of $T_{\rm sub}$.

\section{Conclusions}
\label{sec:conclusion}

We have presented a general, frequency-domain approach to calculate the moments and the (1-point) distribution of the observed Fourier modes characterizing GW signals arising from the superposition of a large number of discrete sources.

Unlike the time-domain treatment of \cite{Ginatetal2020} which focused on bright mergers above a detection threshold, we included in a first step all the GW sources regardless of their evolutionary stage and of the detector sensitivity. Our formulation properly takes into account the observation time $T$ of the experiment, since it controls the convergence to Gaussian distributions along with the number of GW sources overlapping at the detector. Furthermore, the total number of sources $N_0$ which have formed in the observable Universe is also explicit and constrained by the merger rate inferred from data. We illustrated these aspects with a toy model that had only a limited parameter range. We showed that the standard expression of the energy spectrum is recovered, although we emphasize that the source number density on the past-light cone of the observer generally is frequency-dependent owing to the dynamical evolution between the formation of the compact binary and its coalescence.

In addition, we showed analytically that the unsubtracted signal is characterized by a universal $\tilde{h}_f^{-4}$ power-law asymptotic at large strains, where only the coefficient depends on the astrophysical model, at all frequencies; this agrees with the time-domain conclusion of \citet{Ginatetal2020}. This power-law tail is produced by bright, close events.

\begin{figure}
    \centering
    \includegraphics[width=0.45\textwidth]{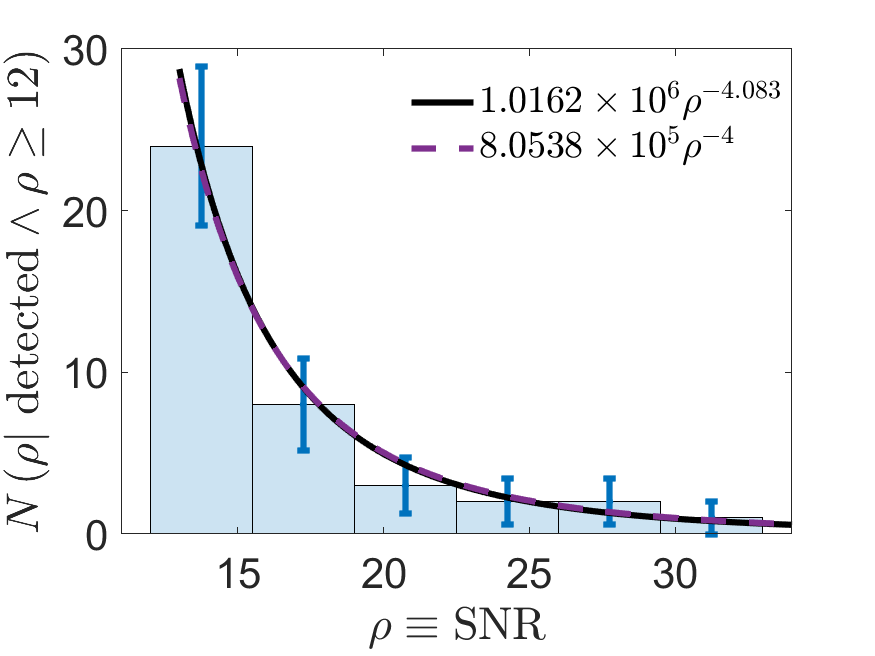}
    \caption{A histogram of the confirmed {\small LIGO}-Virgo-{\small KAGRA} confirmed gravitational-wave events, with SNR $\rho \geq 12$, as a function of the SNR. The number of bins is the maximum number such that no bin is empty. Error bars represent Poisson errors. The black curve is a power-law fit to the histogram, with points weighted by the number of events in each bin ($95\%$ confidence limits of the exponent for this fit are $-4.083 \pm 0.936$), and the purple (dashed) curve is a similar fit, but with the exponent constrained to $-4$.}
    \label{fig:p of snr}
\end{figure}

A simple way to test the $h^{-4}$ prediction with experimental data is to look at all the observed events (the bright sources) and check how their SNR is distributed because it’s essentially proportional to $h$. We tested this with the GWTC catalog of confirmed events from O1-O3.\footnote{\url{https://gwosc.org/eventapi/html/GWTC/}, retrieved 13th May 2023. See also \cite{LIGOvirgo2019PRX,LIGOvirgo2021PRX,LVK2023}.} We only considered events with SNR $\geq 12$, to ensure a complete sample (leaving 40 events), and plotted a histogram, fitting it with a power-law, weighted by the relative frequency to account for the Poissonian errors. The result is presented in figure \ref{fig:p of snr}, which shows good agreement between the observational histogram -- which is proportional to the probability to have an event with SNR $=\rho$ -- and the theoretical prediction of a power-law $\propto \rho^{-4}$.

Our results are valid for cosmological as well as more ``local'' GW signals (such as that produced by galactic white dwarfs). As an illustration, we applied it to the frequency distribution of the unsubtracted signal originating from all the compact binary coalescences (black holes and neutron stars) in the Universe.
Poisson distributed sources and adiabatic sequences of circular orbits evolving under GW emission were assumed for simplicity, but these could be relaxed. The merger rate inferred from the O3 run of the {\small LIGO}-Virgo-{\small KAGRA} collaboration yielded $N_0\sim 10^{18}$, which implied that the $\mathcal{O}(10^{12})$ galaxies of the observable Universe host $\sim 10^6$ active sources on average. As a result, the Fourier modes of the unsubtracted GW signal are close to Gaussian (for an observation time of one year) except for the power-law tail produced by bright mergers.

In a final step, we have applied our approach to quantify deviations from Gaussianity in the resutling SGWB obtained after subtracting bright mergers from the data. 
We have assumed that the source parameters of the bright sources are perfectly known although, in practice, there are uncertainties leading to an additional noise component \citep[see, e.g.][]{Sachdev:2020bkk,Zhou:2022nmt}.
For an observation time of one year and an experiment with A+ sensitivity, the confusion noise produced by unresolved sources ($\textrm{SNR}<12$) is essentially Gaussian for the frequencies considered here. Only much shorter observation times can lead to a retention of the power-law regime, which is then truncated exponentially above a threshold strain. Our results should also be useful for the characterization of confusion noises and for data mining.

\acknowledgments{We would like to thank Matthias Bartelmann and Joseph Allingham for helpful discussions about lensing. Y.B.G., I.R. and V.D. acknowledge funding from the Israel Science Foundation (grant no. 2562/20). Y. B. G. acknowledges support by the Adams Fellowship Programme of the Israeli Academy of Sciences and Humanities. R.R. acknowledges support by the European Research Council (grant number 770935).

This research has made use of data or software obtained from the Gravitational Wave Open Science Center (gwosc.org), a service of LIGO Laboratory, the LIGO Scientific Collaboration, the Virgo Collaboration, and KAGRA. LIGO Laboratory and Advanced LIGO are funded by the United States National Science Foundation (NSF) as well as the Science and Technology Facilities Council (STFC) of the United Kingdom, the Max-Planck-Society (MPS), and the State of Niedersachsen/Germany for support of the construction of Advanced LIGO and construction and operation of the GEO600 detector. Additional support for Advanced LIGO was provided by the Australian Research Council. Virgo is funded, through the European Gravitational Observatory (EGO), by the French Centre National de Recherche Scientifique (CNRS), the Italian Istituto Nazionale di Fisica Nucleare (INFN) and the Dutch Nikhef, with contributions by institutions from Belgium, Germany, Greece, Hungary, Ireland, Japan, Monaco, Poland, Portugal, Spain. KAGRA is supported by Ministry of Education, Culture, Sports, Science and Technology (MEXT), Japan Society for the Promotion of Science (JSPS) in Japan; National Research Foundation (NRF) and Ministry of Science and ICT (MSIT) in Korea; Academia Sinica (AS) and National Science and Technology Council (NSTC) in Taiwan.}

\appendix

\section{Compact binary formation model}
\label{app:galaxymodel}

We summarize our fiducial model here, which, like \cite{Ginatetal2020}, closely follows the `reference model' of \citet{Cusinetal2019}. The inclusion of lensing is separately discussed in Appendix \S\ref{app:lensing}.

The source parameter vector $\vxi=(t_*,a,\imath,M_c,M_\text{gal})$ consists of the cosmic time $t_*$ of binary formation, its initial semi-major axis $a$, orbital inclination $\imath$, chirp mass $M_c$ and the mass $M_\text{gal}$ of the galaxy in which it resides. We approximate the joint probability density $\phi(\vxi)$ by the product
\begin{align}
  \phi(\vxi) &= V_0^{-1}\,\phi(t_*) \,\phi(M_\text{gal}|t_*)\,\phi(m_1,m_2|t_*,M_\text{gal})\nonumber \\ &\qquad \times \,\frac{\phi(\imath)}{a\ln\left(a_{\max}/a_{\min}\right)} \;,
\end{align}
where $V_0$ is the co-moving volume of the observable Universe. This ensures that $\int\!\dd^3\vr\int\!\dd\vxi\,\phi(\vxi) = 1$.
The total number of sources is encapsulated in the value of $N_0$, which is constrained by the observed, present-day merger rate $R_\text{merger}$ of compact stellar remnants (see \S\ref{sec:rates}).

The PDF $\phi(t_*)$ encodes the time dependence of compact binary formation. Neglecting the delay between the formation of stellar and degenerate binaries (which is of order a few $10^6\yr$), we use the cosmic star formation rate and parameterize the distribution $\phi(z_*)$ of binary formation redshift $z_*$ as
\begin{equation}
  \phi(z_*)\,\dd z_* = \frac{(1+z_*)}{\left(\sigma^2+\sqrt{\frac{\pi}{2}} \sigma\right)}\,\mathrm{e}^{-z_*^2/(2\sigma^2)}\,\dd z_*
\end{equation}
Choosing $\sigma = \sqrt{6}$ implies that $\phi(z_*)$ peaks at redshift $z_* = 2$ \cite{Smitetal2012}. $\phi(z_*)$ is eventually converted into a probability distribution $\phi(t_*)$ (per unit cosmic time) using the redshift-to-cosmic-time relation.

To model the mass distribution $\phi(m_1,m_2|t_*,M_\text{gal})$, the initial masses $m_1$ and $m_2$ of the two binary companions are drawn from broken power-law densities, $\phi(m) = Cm^{-\alpha}$ with $\alpha$ dependent on $m$. We choose a Kroupa mass function \cite{Kroupa2002,Binney} (in the mass-range we consider $\alpha=2.7$).

The GW strain is produced by BHs and NSs which formed in the core collapse of massive stars. Their masses are related to the progenitor masses $m_1$ and $m_2$ by the so-called `initial-to-final mass function' $\mu(m,Z)$, which depends on the metallicity $Z$. We use the delayed model presented in \cite{Freyeretal2012} (masses are measured in solar masses):
\begin{widetext}
\be
  \mu(m,Z) = \begin{cases}
               1.3 , & \mbox{if } m \leq 11 \\
               1.1 + 0.2\mathrm{e}^{(m-11)/4} - (2 + Z)\mathrm{e}^{2(m-26)/5}, & \mbox{if } 11 < m \leq 30 \\
               \min\set{33.35 + (4.75 + 1.25Z)(m-34),m-\sqrt{Z}(1.3m-18.35)}, & \mbox{otherwise}
             \end{cases}
\ee
\end{widetext}
The metallicity depends on the cosmic time of formation, a dependence which we model (following again \cite{Cusinetal2019}) using the fit of \cite{Maetal2016}, \emph{viz.}
\begin{align}
\label{eqn:metallicity as function of redshift}
  \log_{10}\left(\frac{Z(z,M_{\textrm{gal}})}{Z_\odot}\right) &= 0.35\left[\log_{10}\left(\frac{M_{\textrm{gal}}}{M_\odot}\right) - 10\right] \nonumber \\ &\qquad + 0.93\mathrm{e}^{-0.43z} - 1.05 \;.
\end{align}
The redshift $z$ is converted to cosmic time $t_*$ assuming a $\Lambda$CDM cosmology.

Combining these various relations leads to a mass distribution
\be
\phi(m_1,m_2|t_*,M_\text{gal})\equiv \phi(m_1,m_2|Z(t_*,M_\text{gal})
\ee
given by
\begin{align}
  \phi(m_1,m_2|Z) &= \iint \mathrm{d}\tilde{m}_1\mathrm{d}\tilde{m}_2\, \phi(\tilde{m}_1)\phi(\tilde{m}_2)\\
  &\times\delta^\text{D}\!\left(m_1 - \mu(\tilde{m}_1,Z)\right)\delta^\text{D}\!\left(m_2 - \mu(\tilde{m}_2,Z)\right), \nonumber
\end{align}
where $\delta_\text{D}(x)$ is the Dirac delta-function.

The next ingredient is $\phi(M_\textrm{gal}|t_*)$, which we model using the halo mass function of \citet{Tinkeretal2016}, assuming that the total stellar mass in a galaxy is proportional to its halo mass $M_\text{h}$.

Finally, we assume a uniform distribution $\phi(\imath)$ of orbital inclination, and a $1/a$ scaling for the PDF of the initial semi-major axis, in agreement with \"{O}pik's law \cite{Oepik1924}. The latter approximates the observed Galactic period distribution reasonably, over a fairly large range of periods \cite{DucheneKraus2013}.
The limits $a_{\min} = 0.014~\textrm{AU}$ and $a_{\max} = 4000~\textrm{AU}$ are adopted as in \cite{Cusinetal2019}. They translate into limits on the initial period $T_*$ (by Kepler's third law) using the masses of the binary components.

\begin{figure}
    \centering
    \includegraphics[width = .45\textwidth]{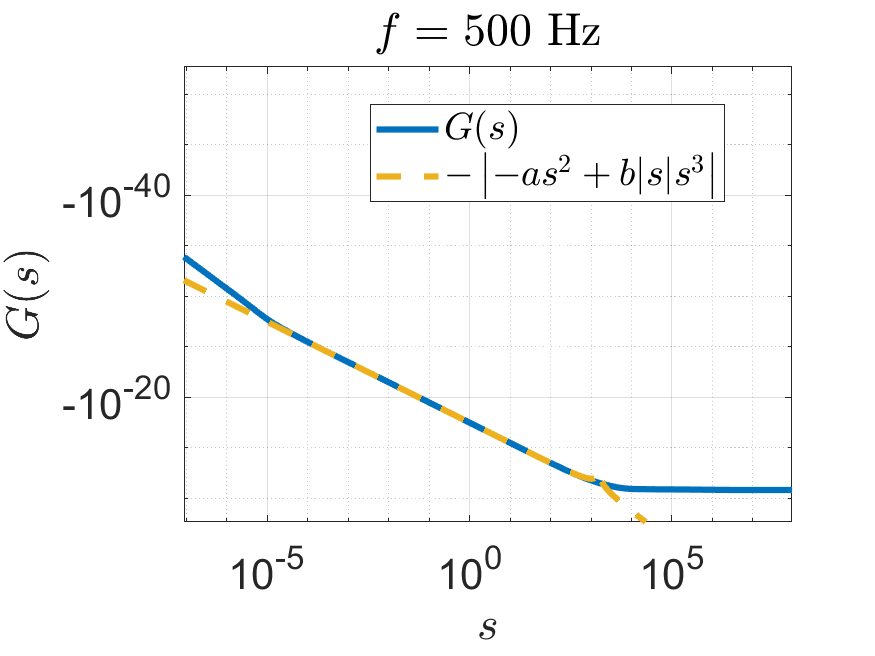}
    \caption{The function $G(s,f)$ (solid curve) as well as its cubic approximation $G(s;f)=-a(f) s^2 + b(f) |s|^3$ (dashed curve), with $a(f)$ and $b(f)$ computed from Eqs~(\ref{eq:af}) and (\ref{eq:bf}).}
    \label{fig:G 100 Hz}
\end{figure}

The characteristic function $G(s;f)$ is obtained from an integration over $\vxi$.  The knowledge of $T_*$ and $M_c$ determines the binary lifetime $\tau_0$. 

\section{Small-$s$ Limit of $G$}
\label{app:G approximation}

In this Appendix, we derive the expressions for the dimensionless coefficients $a(f)$ and $b(f)$ appearing in equation \eqref{eqn:G small s approximation}. As in appendix A of ref. \cite{Ginatetal2020}, these correspond to the poles at $\mu = -2$ and $\mu = -3$ of the Mellin transform $\overline{G}(\mu;f)$, which is given by
\begin{widetext}
\begin{equation}
    \overline{G}(\mu;f) = 2^{\mu-1}\frac{\Gamma\!\left(\frac{\mu}{2}\right)}{\Gamma\!\left(1-\frac{\mu}{2}\right)}
    \int_0^1\!\dd\tau\,\tau^2\int\!\dd^2\nvh \int\! \mathrm{d}\vxi\,\phi(\tau,\nvh,\vxi) \left(\frac{A_f(\tau,\nvh,\vxi)}{\tau}\right)^{-\mu}
    \;,
\end{equation}
\end{widetext}
where $\tau=r/r_0$ is the dimensionless co-moving distance and $A_f(\tau,\nvh,\vxi) \equiv \tau|\tilde h_{f,\nvh}|/h_c$ is the coefficient of $s$ in the Bessel function's argument multiplied by $\tau$. Note that, although $|\tilde h_{f,\nvh}|\propto 1/r$, $A_f$ has a residual, weak dependence on $\tau$ through the source redshift $z=z(r)$. Furthermore, as already mentioned, we restrict ourselves to a spatial Poisson process so that $\phi(\tau,\nvh,\vxi)$ does not explicitly depend on $(\tau,\nvh)$.
The integral
\be
    I(\mu) \equiv \int_0^1\!\dd\tau\,\tau^2\int\!\dd^2\nvh \int\! \dd\vxi\, \phi(\tau,\nvh,\vxi) \left(\frac{A_f(\tau,\nvh,\vxi)}{\tau}\right)^{-\mu}
\ee
is analytic at $\mu = -2$, but has a pole at $\mu = -3$. Therefore the residue of $\overline{G}(\mu;f)$ at $\mu = -2$ is just
\begin{multline}
I(-2)\,\textrm{Res}\left(2^{\mu-1}\frac{\Gamma\!\left(\frac{\mu}{2}\right)}{\Gamma\!\left(1-\frac{\mu}{2}\right)},\mu = -2\right) \\ = -\frac{1}{4}\,\int_0^1\!\dd\tau\int\!\dd^2\nvh\int\!\dd\vxi\,\phi(\tau,\nvh,\vxi)\,A_f^2(\tau,\nvh,\vxi)\;.
\end{multline}
For $\mu = -3$, it is clear that the pole comes from $I(\mu)$ itself. Therefore, we may write
\begin{align}
    I(\mu) &= \int_0^1\!\dd\tau\, \tau^2\int\!\dd^2\nvh \int\!\dd\vxi \,\phi(0,\nvh,\vxi) \left(\frac{A_f(0,\nvh,\vxi)}{\tau}\right)^{-\mu}
    \nonumber \\
    &\qquad + \mbox{analytic at }\mu = -3 \label{eq:Imu1} \;,
\end{align}
because both $A$ and $\phi$ are analytic at $\tau = 0$. Since the behavior of $A$ and $\phi$ away from $\tau = 0$ is immaterial for the residue at $\mu=-3$, we may use instead
\begin{align}
    I(\mu) &= \int_0^{\infty}\! \dd\tau\, \tau^2\int\!\dd^2\nvh \int\!\dd\vxi \,\phi(0,\nvh,\vxi) \left[\frac{A_f(0,\nvh,\vxi)}{\tau\,\mathrm{e}^{\tau}}\right]^{-\mu} \nonumber \\
    &\qquad + \mbox{analytic at }\mu = -3
\end{align}
to compute it, again, because $\mathrm{e}^{-\tau\mu}$ is analytic. This expression differs from (\ref{eq:Imu1}) only by an analytic function. Performing the integral over $\tau$ gives
\begin{align}
    I(\mu) &= \frac{\Gamma\!\big(3+\mu\big)}{(-\mu)^{3+\mu}}\int\!\dd^2\nvh \int\!\dd\vxi \,\phi(0,\nvh,\vxi)\,\Big[A_f(0,\nvh,\vxi)\Big]^{-\mu}\nonumber \\
    &\qquad + \mbox{analytic at }\mu = -3 \label{eq:Imu2} \;,
\end{align}
so that the residue of $\bar G(\mu;f)$ at $\mu=-3$ is
\begin{multline}
2^{\mu-1}\frac{\Gamma\!\left(\frac{\mu}{2}\right)}{\Gamma\!\left(1-\frac{\mu}{2}\right)}\, \textrm{Res}\Big(I(\mu),\mu=-3\Big) \\
= \frac{1}{9}\int\!\dd^2\nvh\int\!\dd\vxi\,\phi(0,\nvh,\vxi)\,A_f^3(0,\nvh,\vxi) \;.
\end{multline}
One may now proceed as in Ref.~\cite{Ginatetal2020} to find the small-$s$ expansion $G(s;f)\approx -a(f)s^2 + b(f) |s|^3$, with
\be
    \label{eq:af}
    a(f)= \frac{1}{4}\int_0^1\!\dd\tau\int\!\dd^2\nvh\int\!\dd\vxi\,\phi(\tau,\nvh,\vxi)\,A_f^2(\tau,\nvh,\vxi)
\ee
and
\be
    \label{eq:bf}
    b(f) = \frac{1}{9}\int\!\dd^2\nvh\int\!\dd\vxi\,\phi(0,\nvh,\vxi)\, A_f^3(0,\nvh,\vxi)
\ee
Observe that $a(f)$ is equal to
\be
a(f)=\frac{1}{4 h_c^2}\,\mathbb{E}\big[|\tilde h_{f,\nvh}|^2\big] \;,
\ee
which ensures that we recover $\langle |\tilde h_{f,\nvh}|^2\rangle = N_0\, \mathbb{E}\big[|\tilde h_{f,\nvh}|^2\big] = \frac{F}{2}S_h(f)$ for the second moment of the distribution $P(|\tilde h_{f,\nvh}|)$ of the DFT amplitudes.

For illustration, we plot $G(s;f)$ in Fig.~\ref{fig:G 100 Hz} for the full model described in Appendix \S\ref{app:galaxymodel} and an observed frequency $f = 500\Hz$.
The approximation \eqref{eqn:G small s approximation} remains valid at small $s$, where the numerical integration becomes quite noisy, and for such $s$ we use the asymptotics in evaluating the Hankel transform.

\section{Comparison with previous literature}
\label{appendix:comparison with Phinney}

Refs.~\cite{phinney:2001,schneider/ferrari/etal:2001} outline a simple approach based on time-domain GW strain fluctuations to calculate the energy spectrum of any SGWB, which is widely used in the literature \cite[e.g.][]{sesana/etal:2008,Regimbau:2011rp,rosado:2011,marassi/schneider/etal:2011,zhu/howell/etal:2011,wu/mandic/regimbau:2012,koko/etal:2015,AbbottSGWB2018,capurri/etal:2021}. Let us check whether we recover their expression for $\Omega_\text{gw}(f)$.

We start from Eq.~(\ref{eq:ToyOmegagw1}) and insert the square of the amplitude $h_0(f)$ given by Eq.~(\ref{eq:gwamplitude}). Using the expression
\be
    \frac{\dd E_s}{\dd\ln f_s} = \frac{\pi^{2/3}}{3G}(GM_c)^{5/3}f_s^{2/3}
\ee
for the (rest-frame) energy $\dd E_s$ emitted by a single compact binary in the (rest-frame) frequency interval $\dd\ln f_s$, we can recast Eq.~(\ref{eq:ToyOmegagw1}) into the functional form adopted in \cite{phinney:2001},
\be
    \label{eq:ToyOmegagw2}
    \Omega_\text{gw}(f) = \frac{1}{\rho_c}
    \int_0^{r_0}\!\dd r\int\!\dd M_c\,\frac{\partial^2 n}{\partial r\partial M_c}\,\frac{1}{(1+z)}\,\frac{\dd E_s}{\dd\ln f_s}\;,
\ee
after substituting $d_L(z) = (1+z) r$. In Eq.~(\ref{eq:ToyOmegagw2}), the factor of $(1+z)^{-1}$ takes into account the redshift of gravitons due to the expansion of the Universe whereas, in the co-moving number density of sources per unit radial co-moving distance and chirp mass,
\begin{align}
    \frac{\partial^2 n}{\partial r\partial M_c} &= \frac{1}{c}\left(\frac{N_0}{V_0}\right)(1+z)^{-1}\,\phi\big(t_*(f,\eta_0,r,M_c)\big)\,\phi(M_c)
    \nonumber \\
    &\equiv \frac{1}{c}\,(1+z)^{-1}\,R_*(f,\eta_0,r,M_c)\;,
    \label{eq:dn2drdm}
\end{align}
the factor of $(1+z)^{-1}$ converts the infinitesimal physical separation $c\, \dd t_*$ measured in the source rest frame (along the propagation direction of the GW signal) into a co-moving separation $\dd r$. The second equality rephrases the source number density in terms of a binary event rate $R_*(f,\eta_0,r,M_c)$ per co-moving volume and chirp mass. Note that, owing to the dynamical evolution from binary formation until coalescence, both $\partial^2 n/\partial r\partial M_c$ and $R_*$ depend on the observed frequency via $t_*(f,\eta_0,r,M_c)$. However, this dependence is very mild because
\be
    \frac{\dd\ln t_*}{\dd\ln f} = -\frac{8}{3(1+y)} \;,\quad\mbox{with}\quad y=\frac{t_{0,\ret}-\tau_0}{t_*} \;,
\ee
and the dimensionless parameter $y$ is $|y|\ll 1$ for most binaries (since $\tau_0\gg t_0$ typically) unless the compact binary is about to merge.

Eq.~(\ref{eq:dn2drdm}) provides the connection between our frequency-domain approach and standard computations of GW energy spectra.

\section{Strong gravitational lensing}
\label{app:lensing}

Let us describe the lensing probability model for $\phi(\mu|z)$. We follow the procedure outlined in \cite{Robertsonetal2020}. The cross-section $\sigma_\text{E}$ for a magnification $|\mu|>\mu_0$ for a source at redshift $z=z_s$ is
\be
    \sigma_\text{E}(|\mu|>\mu_0,M_\text{h},z_l,z_s) = \theta_\text{E}^2(M_\text{h},z_l,z_s)\, f(\mu_0)
\ee
where $M_\text{h}$ is the virial mass of the $z_l<z_s$ lens (halo) and
\be
    \theta_\text{E}(M_\text{h},z_l,z_s) = \frac{M_\text{h}}{2\, R_\text{h}\,\Sigma_\text{crit}\, d_A(z_l)}
\ee
is the Einstein radius. The halo virial radius $R_\text{h}$ and virial mass $M_\text{h}$ are defined assuming a standard density threshold $\Delta_c=200$ times the critical background density $\rho_c(z)$, while $d_A$ will designate the angular diameter distance. Furthermore, the function $f(\mu_0)$ encodes the dependence on the mass profile of the lens. We shall adopt the functional form
\be
    f_\text{SIS}(\mu_0) = \pi \begin{cases}
                                \frac{1}{(\mu - 1)^2}, & \mbox{if } 1 < \mu \leq 2\\
                                \frac{4}{\mu^2}, & \mbox{if } 2 < \mu
                              \end{cases}
\ee
corresponding to a singular isothermal sphere (SIS) \citep{Schneideretal1992}.

The critical surface density for lensing is given by
\begin{align}
    \Sigma_\text{crit} &=\left(\frac{2\pi G}{c^2}\right) \frac{M_\text{h} d_A(z_l,z_s)}{R_\text{h} d_A(z_s)} \\ &=\left(\frac{c^2}{4\pi G}\right)\frac{d_A(z_s)}{d_A(z_l) d_A(z_l,z_s)} \nonumber
\end{align}
where $d_A(z_l,z_s)$ is the angular diameter distance between the source and the lens. Putting these relations together gives
\be
    \theta_E \simeq 2.1\times  10^{-14} \left[\frac{M}{M_\odot}\right]^{2/3} \left[\frac{\Delta_c}{200}\right]^{1/3}
    E(z)^{2/3}\, \frac{d_A(z_l,z_s)}{d_A(z_s)}\;.
\ee
Note that regions of the source plane can map to multiple regions in the image plane if the source falls within the Einstein ring.

The total cross-section per unit lens mass and redshift for a source at redshift $z_s>z_l$ is
\begin{multline}
  \frac{\partial^2 \sigma_\text{tot}}{\partial z\partial M}(>\!\mu_0,M_\text{h},z_l,z_s) = \bar n(M_\text{h},z_l) \\ \times \sigma_\text{E}(>\!\mu_0,M_\text{h},z_l,z_s)
  \frac{dV}{dz}(z_l) \;.
\end{multline}
Here, $\bar n(M_\text{h},z)$ be the differential halo mass function and $V(z)$ is the co-moving volume out to redshift $z$.
The total optical depth is (we assume an all-sky survey appropriate to GW detectors)
\begin{align}
    \tau(>\!\mu_0,z_s)
    = & \frac{1}{4\pi}\;\int_0^{z_s}\!\dd z\,\int\!\dd^2\nvh\,\int \!\dd M\,\nonumber\\
    & \times\frac{\partial^2 \sigma_\text{tot}}{\partial z\partial M}(>\!\mu_0,M,z,z_s) \nonumber \\
    = &\; \int_0^{z_s}\!\dd z \,r^2 \frac{\dd r}{\dd z} \int\!\dd M\,
    \bar n(M,z)\,\nonumber\\
    &\;\times\sigma_\text{E}(>\!\mu_0,M,z,z_s) \nonumber \;.
\end{align}
For $\mu_0=2$, $\tau(>\!\mu_0,z_s)$ does not exceed $\mathcal{O}(10^{-2})$ even for redshifts as large as $z=5$.

The probability for an image of a source at redshift $z_s$ being lensed by more that $\mu_0$ is simply $\tau$, in the linear regime (if $\tau$ is small) \citep{Schneideretal1992}, but is otherwise given by a non-linear functional of $\tau$ \citep{Pei1993a,Pei1993b}, which is beyond the scope of this work. One therefore may approximate the lensing cumulative probability by \citep[chapter 12]{Schneideretal1992}
\begin{equation}\label{eqn:lensing P discontinuous}
  P_1(> \mu|z_s) = \begin{cases}
                   \tau(\mu,z_s), & \mbox{if } \mu > \mu_L(z_s) \\
                   1, & \mbox{if } 1 \leq \mu \leq \mu_L(z_s)
                 \end{cases},
\end{equation}
where $\mu_L(z_s)$ is some cut-off magnification at which $\tau(\mu_L(z_s),z_s) \lesssim 1$. In practice, it is preferable to adopt a differentiable probability density $\phi$, which approximates the above equation. For this purpose, we follow a procedure not dissimilar to, e.g., \cite{BartelmannSchneider1990}, and assume the following probability density function
\be
  \phi(\mu|z_s) = \begin{cases}
                 \frac{\sqrt{2}\alpha(z_s)}{\sqrt{\pi}\mu \sigma(z_s)}\mathrm{e}^{-\ln^2\mu/(2\sigma^2(z_s)}, & \mbox{if } 0 < \mu < 1 \\
                   \frac{1-\alpha(z_s)}{\lambda(z_s)}\mathrm{e}^{-(\mu - 1)/\lambda(z_s)}, & \mbox{if } 1 \leq \mu \leq \mu_L(z_s) \\
                   (\alpha(z_s)-1)\frac{\partial\tau}{\partial \mu}, & \mbox{if } \mu > \mu_L(z_s).
                 \end{cases}
\ee
This is a Gaussian in $\ln \mu$ for $\mu < 1$, decays exponentially until it becomes linear for small magnifications, and becomes $(\alpha(z_s)-1)\frac{\partial\tau}{\partial \mu}$ at large $\mu$. The parameters (functions of source redshift) $\alpha$, $\sigma$, $\lambda$, $\mu_L$ are fixed by
\begin{enumerate}
    \item continuity at $\mu = 1$,
    \item a mean magnification $\langle \mu \rangle = 1$ (as appropriate for the kind of cosmological distribution of lenses and sources we study here \cite{KaiserPeacock2016}),
    \item continuity at $\mu = \mu_L$ at all redshifts,
    \item and normalization $\phi(>0|z_s) = 1$, for all $z_s$.
\end{enumerate}
Continuity at $\mu_L$ is ensured by requiring
\be\label{eqn:lensing continuity condition}
  \frac{1}{\lambda}\exp\left[\frac{\mu_L - 1}{\lambda}\right] = -\tau'(\mu_L).
\ee
The normalization $\phi(\mu > 0|z_s) = 1$ for all $z_s \geq 0$ constrains
\be\label{eqn:lensing normalisation condition}
  \frac{\mu_L(z_s) - 1}{\lambda(z_s)} = \ln \frac{1}{\tau( > \mu_L,z_s)}\;.
\ee
For an SIS, $\tau(> \mu,z_s) \equiv f_{\rm SIS}(\mu)g(z_s)$, whence for $1 < \mu_L < 2$ condition \eqref{eqn:lensing continuity condition} becomes
\be
  \frac{\mu_L - 1}{\lambda} + \ln \left[\pi g(z_s)\right] = 2 \ln (\mu_L - 1).
\ee
Upon defining $y \equiv \frac{\mu_L - 1}{\sqrt{\pi g}}$, $u = \frac{\lambda}{\sqrt{\pi g}}$, conditions \eqref{eqn:lensing continuity condition} and \eqref{eqn:lensing normalisation condition} become
\begin{align}
  & u = \frac{y}{2\ln y} \\ &
  \frac{y}{u}\mathrm{e}^{-y/u} = \frac{2}{y^2},
\end{align}
which are solved by $y = \mathrm{e}$ and $u = \mathrm{e}/2$, or
\begin{align}
  \mu_L(z_s) & = 1 + \mathrm{e}\sqrt{\pi g(z_s)} \\
  \lambda(z_s) & = \frac{\mathrm{e}}{2}\sqrt{\pi g(z_s)}.
\end{align}

Continuity at $\mu = 1$ is tantamount to setting
\be\label{eqn: lensing appendix continuity at 1}
    \frac{\sqrt{2}\alpha}{\sqrt{\pi} \sigma} = \frac{(1-\alpha)}{\lambda},
\ee
and the expectation value condition is satisfied when
\begin{widetext}
\be
    \alpha\left[1-\mathrm{e}^{\sigma^2/2}\mathrm{erfc}\left(\frac{\sigma}{\sqrt{2}}\right)\right] = (1-\alpha)\left[\frac{1}{2} \left(\frac{\sqrt{\pi g} \left(2 \sqrt{\pi g} \mu_L^2+\mathrm{e} (\mu_L-1)^2\right)}{(\mu_L-1)^2}-\mathrm{e}^{-\frac{2 (\mu_L-1)}{\mathrm{e} \sqrt{\pi g}}} \left(\mathrm{e} \sqrt{\pi g}+2 \mu_L\right)\right)\right].
\ee
Dividing the two above equations by each other eliminates $\alpha$:
\be
    \sqrt{\frac{\pi}{2}}\sigma\left[1-\mathrm{e}^{\sigma^2/2}\mathrm{erfc}\left(\frac{\sigma}{\sqrt{2}}\right)\right] = \lambda\left[\frac{1}{2} \left(\frac{\sqrt{\pi g} \left(2 \sqrt{\pi g} \mu_L^2+\mathrm{e} (\mu_L-1)^2\right)}{(\mu_L-1)^2}-\mathrm{e}^{-\frac{2 (\mu_L-1)}{\mathrm{e} \sqrt{\pi g}}} \left(\mathrm{e} \sqrt{\pi g}+2 \mu_L\right)\right)\right].
\ee
As $g \ll 1$ at all redshifts, and consequently so are $\lambda$ and $\sigma$, we expand the error function, and approximate the solution to this equation by
\be
    \sigma \approx \left[\left((1+\mathrm{e}^2)\frac{\sqrt{\pi g}}{2\mathrm{e}} + \pi g\right)\frac{\mathrm{e}\sqrt{\pi g}}{2}\right]^{2/3}.
\ee
We use this in computing $G$, rather than the exact, numerical solution, to have an analytic $\phi(\mu|z_s)$ which can be quickly evaluated. For the relevant range of optical depth ($g\lesssim 0.01$), the relative error between this solution and the exact solution is a few percent (which is likely smaller than the error introduced upon modeling all the lenses as SISs).

From \eqref{eqn: lensing appendix continuity at 1}, we have exactly
\be
    \alpha = \frac{1}{\frac{\lambda}{\sigma}\sqrt{\frac{2}{\pi}}+1},
\ee
which ensures that $0 \leq \alpha \leq 1$, and hence that the probability distribution is normalized to unity.

The full lensing probability distribution function is thus given by
\be\label{eqn:lensing p continuous}
  \phi(\mu|z_s) = \begin{cases}
                 \frac{\sqrt{2}\alpha(z_s)}{\sqrt{\pi}\mu \sigma(z_s)}\mathrm{e}^{-\ln^2\mu/(2\sigma^2(z_s)}, & \mbox{if } 0 < \mu < 1 \\
                 \frac{2\left[1-\alpha(z_s)\right]}{\sqrt{\pi g(z_s)}}\exp\left[\frac{2(1-\mu)}{\mathrm{e}\sqrt{\pi g(z_s)}} - 1\right], & \mbox{if } 1 \leq \mu \leq \mu_L(z_s) \\
                 \frac{2\pi g(z_s)\left[1-\alpha(z_s)\right]}{(\mu - 1)^3}, & \mbox{if } \mu_L(z_s) < \mu \leq 2 \\
                 \frac{8\pi g(z_s)\left[1-\alpha(z_s)\right]}{\mu^3}, & \mbox{if } 2 < \mu \\
                 0, & \mbox{otherwise}.
               \end{cases}
\ee
This $\phi(\mu|z_s)$ satisfies conditions $1,3$ and $4$ exactly, and $2$ to within $6.3\%$ at redshift $10$ (worst case -- at redshift $1$, $\langle \mu \rangle = 1.02$). The jump discontinuity at $\mu = 2$ is a property of the SIS, where a second image appears at total magnification $\mu = 2$ \cite{Schneideretal1992}.
\end{widetext}

\begin{figure}
  \centering
  \includegraphics[width=0.45\textwidth]{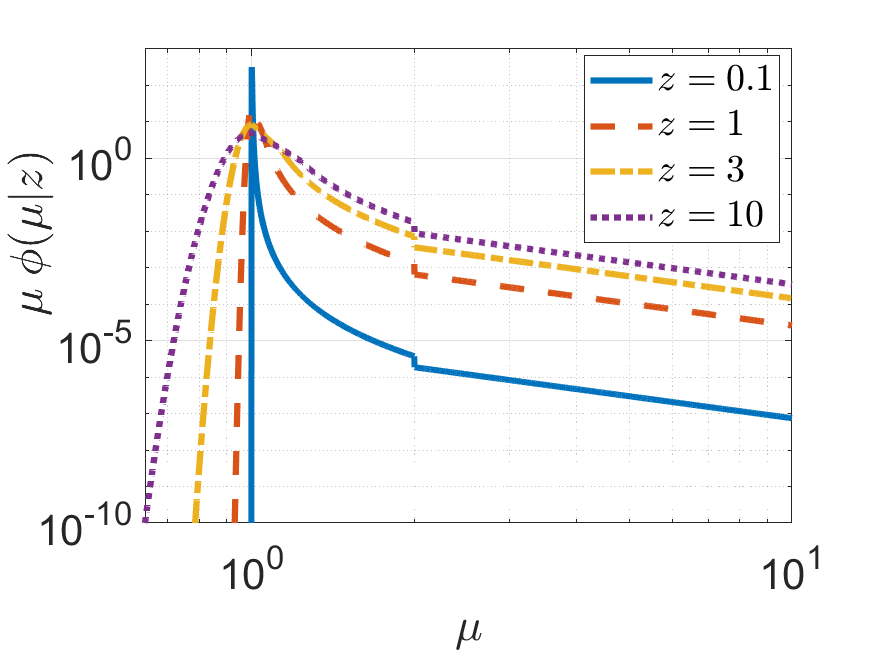}
  \caption{The magnification probability density $\mu\phi(\mu|z)$ obtained from Eq.~\eqref{eqn:lensing p continuous} for selected values of the source redshift. \label{fig:magnificationPDF}}
\end{figure}

We plot $\phi(\mu|z_s)$ for various redshifts in figure \ref{fig:magnificationPDF}. Since $g(z) \to 0$ as $z \to 0$, so do $\lambda(z_s)$ and $\sigma(z_s)$, and we have
\begin{equation}
  \phi(\mu|z_s) \underset{z_s \to 0}{\longrightarrow} \delta^{\rm D}(\mu - 1),
\end{equation}
as it should, physically \citep{Pei1993a}, because there is no lensing for a source at the observer's position.

In the evaluation of $\phi(\mu|z_s)$ at very low redshifts, numerical errors in exponentials of very large, negative numbers sometime lead the computer to erroneously set $\phi(\mu|z) = 0$ for $\mu < 1$, which leads to an un-normalized probability distribution, because $\alpha < 1$. We solve this in practice with the replacement $\phi(\mu|z_s) \mapsto \phi(\mu|z_s)/[1-\alpha(z_s)]$ for $\mu \geq 1$, whenever the computer evaluates $\phi(\mu = 0.99|z_s) = 0$.

\section{Bright Source Subtraction}
\label{app:bright sources}

We wish to approximate
\be\label{eqn:appendix P_r definition}
    P_r(h) = h \int_0^\infty \mathrm{d}q ~q J_0(qh)\mathrm{e}^{N_0G(q)} \;,
\ee
at $h\to \infty$, where $G(q)$ is now an analytic function. We use the same approach as in appendix C of \cite{Ginatetal2020} -- the method of steepest descents.
Before proceeding, note that we may replace the $J_0$ in \eqref{eqn:appendix P_r definition} by a Hankel function $H_0^{(1)}(hq)$, \emph{viz.}
\be
    P_r(h) = \frac{h}{2}\int_{-\infty}^{\infty}\!\dd q\, q\, H_0^{(1)}(hq) \mathrm{e}^{N_0G(q)}\;.
\ee
One can write
\be
    H_0^{(1)}(z) = \sqrt{\frac{2}{\pi z}}\mathrm{e}^{\mathrm{i}(z-\pi/4)}\left[\sum_{m=0}^{p-1}\frac{\mathrm{i}^m a_m(0)}{z^m} + O(z^{-p})\right],
\ee
{which is true for $-\pi /2 < \arg z < 3\pi/2$, \cite[p. 219]{Watson1944}, and $a_k(\nu)$ is given by equation \eqref{eqn:a k of nu}, and we choose the branch cut along the negative imaginary axis. Then the exponent becomes $\mathrm{i}hq + N_0G(q)$.}
The exponent has a stationary point when
\be
\label{eqn:appendix_bright_source_stationary_point_condition}
    \mathrm{i}h = N_0G'(s)h_c \;.
\ee
Na\"{i}vely, it might seem that the large $h$ tail of $P_r(h)$ stems from the small $s$ limit of $G(s)$. But we know that at small $s$, $G(s) \sim -as^2$, and if $s \ll 1$, this is solved by $s = -\mathrm{i}h/(2N_0ah_c)$ which might be small for intermediate values of $h$, but isn't for arbitrarily large $h$. For such intermediate values of $h$, this implies that the Hankel transform is dominated by the second moment of $G(s)$, i.e. by the Gaussian part.

Therefore, the solution to equation \eqref{eqn:appendix_bright_source_stationary_point_condition} may only emerge at large $s$,
if one is interested in sufficiently large $h$. For real $s$, the left-hand side is of order $h$, while the right-hand side remains bounded for any $s$. The solution therefore lies in complex values of $s$. If we shift the integration contour (up or down) and substitute $q = x \pm \mathrm{i}y$, then the $J_0$ in the integrand of $G$ becomes unbounded, which may lead to a possible increase in the right-hand side, and hence allows for a solution when $h$ is large. For exactly the same reasons as in \cite{Ginatetal2020}, the analytic continuation of $G$ to complex, large, values of $s$, is
\be
    G(x \pm\mathrm{i}y) \sim C\frac{\mathrm{e}^{\mp \mathrm{i}h_{\max}q \pm \mathrm{i}\pi/4}}{q^{3/2}} \;,
\ee
where $h_{\rm max}$ is the maximum value of $h_0/\sqrt{T}$ that satisfies the SNR condition with an equality,
\be
    C = \int\mathrm{d}\boldsymbol{\xi} \left. \frac{h_c\sqrt{T}}{\sqrt{2 \pi}}\abs{\frac{\partial r}{\partial \tilde h}} \phi(\boldsymbol{\xi})\, \Pi_T\, \left( \frac{h_c}{h_{\rm max}}\right)^{1/2} \right|_{r=\chi(h_{\rm max},\boldsymbol{\xi})}
\ee
(recall that, while $h_c$ and $h_\text{max}$ have units of $[{\rm time}]^{1/2}$, $\tilde h$ has units of time) and
\be
    \chi(h_0,\boldsymbol{\xi}) \equiv \frac{1}{\pi^{2/3}}\sqrt{\frac{5}{24}}\frac{a c}{h_0}\left(\frac{GM_c}{ac^3}\right)^{5/6} A_{\rm GR} \;.
\ee
This approximation for $G(x\pm\mathrm{i}y)$ follows from an application of Laplace's method, and the approximation $J_0(z) \sim \cos (z-\pi/4)\sqrt{2/(\pi z)}$ as $\abs{z} \to \infty$ \cite{DLMF}.

The derivative becomes (to leading order)
\be
    G'(q) \sim \mp \mathrm{i}h_{\max} G(q) \;.
\ee

Using $H_0^{(1)}(z) \sim \sqrt{\frac{2}{\pi z}}e^{\mathrm{i}z - \mathrm{i}\pi/4}$ for large $\abs{z}$, yields an exponent
\be
    \zeta(q) \equiv \mathrm{i}h q + N_0C\frac{\mathrm{e}^{- \mathrm{i}h_{\max}q + \mathrm{i}\pi/4}}{q^{3/2}} \;,
\ee
where we chose the positive sign, to comply with the Hankel function's approximation validity regime. The stationary point condition $\zeta'(q) = 0$ yields
\be
    \frac{h}{h_{\max}} = -N_0C\frac{\mathrm{e}^{-\mathrm{i}h_{\max}q + \mathrm{i}\pi/4}}{q^{3/2}} \;.
\ee
Equating the modulus and phase implies that at the stationary point
\begin{align}
    h_{\max} x_{sp} & = \pi\left(\frac{1}{2}+2k\right), \\
    h_{\rm max} y_{sp} & = -\frac{3}{2}W_{-1}\!\!\left[-\frac{2h_{\max}}{3}\left(\frac{N_0h_{\max}C}{h}\right)^{2/3}\right]\;,
\end{align}
where $k$ is an integer and $W_{-1}$ is the secondary branch of Lambert's $W$-function, and we have approximated $\abs{q}^{3/2} \approx y^{3/2}$, and $\arg (x\pm \mathrm{i}y) \approx \pm \pi/2$, because $h_{\max} y_{sp} \sim \ln h/h_{\max} \gg h_{\max} x_{sp}$, and we only consider the $k=0$ saddle because this will have the dominant contribution.

\begin{widetext}
Hence,
\begin{align}
    \mathrm{i}h(x_{sp} + \mathrm{i}y_{sp}) & = \mathrm{i}h_{\max}\frac{h(x_{sp}+\mathrm{i}y_{sp})}{h_{\max}} = -\frac{N_0C}{\sqrt{x_{sp}^2 + y_{sp}^2}} \exp\left(h_{\max} y_{sp} - \mathrm{i}h_{\max}x_{sp} +\mathrm{i}\frac{\pi}{4} - \mathrm{i}\frac{\arg (x_{sp} + \mathrm{i}y_{sp})}{2} + \mathrm{i}\frac{\pi}{2}\right) \nonumber \\ &
    = -\frac{N_0C}{\sqrt{x_{sp}^2 + y_{sp}^2}} \exp\left(h_{\max} y_{sp}\right) \approx -\frac{3h}{2h_{\max}}\abs{W_{-1}\left[-\frac{2h_{\max}}{3}\left(\frac{N_0h_{\max}C}{h}\right)^{2/3}\right]} \;.
\end{align}
The exponent becomes
\be
    \zeta(s) \approx -\frac{h}{h_{\max}} -\frac{3h}{2h_{\max}}\abs{W_{-1}\left[-\frac{2h_{\max}}{3}\left(\frac{N_0h_{\max}C}{h}\right)^{2/3}\right]} - \frac{h h_{\max}}{2}(s-x_{sp} -\mathrm{i}y_{sp})^2 + \ldots
\ee
Since the coefficient of $(s-x_{sp} -\mathrm{i}y_{sp})^2$ is negative, the steepest descent contour is parallel to the real axis, with $y=y_{sp}$.

The integral becomes
\begin{align}
    P_r(h) & \sim \frac{h}{\sqrt{2\pi}}\int_{-\infty + \mathrm{i}y_{sp}}^{\infty + \mathrm{i}y_{sp}} \sqrt{s} \mathrm{e}^{\zeta(s)-\mathrm{i}\pi/4} \mathrm{d}s = \frac{h}{\sqrt{2\pi}}\int_{-\infty}^{\infty} \left(\frac{\pi^2}{4h_{\max}^2} + y_{sp}^2\right)^{1/2}\mathrm{e}^{\mathrm{i}\arg s/2} \mathrm{e}^{\zeta(x+\mathrm{i}y_{sp})-\mathrm{i}\pi/4} \mathrm{d}x \\ &
    \sim \frac{h}{\sqrt{2\pi}}\int_{-\infty}^{\infty} \mathrm{e}^{\zeta(x+\mathrm{i}y_{sp})}\left(\frac{\pi^2}{4h_{\max}^2} + y_{sp}^2\right)^{1/2}  \mathrm{d}x \nonumber \\ &
    \sim \sqrt{\frac{h}{h_{\max}^3}} \exp\left(-\frac{h}{h_{\max}} -\frac{3h}{2h_{\max}}\abs{W_{-1}\left[-\frac{2h_{\max}}{3}\left(\frac{N_0h_{\max}C}{h}\right)^{2/3}\right]}\right)\left(\frac{\pi^2}{4} + \ln^2\frac{h}{h_{\max}}\right)^{1/2} \;.\nonumber
\end{align}
This is the shape of the probability distribution of the confusion background in the limit of large $h_\text{max}$ and $h>h_\text{max}$.
\end{widetext}

\bibliography{sgwb}

\end{document}